\documentclass[aps,prx,twocolumn,showpacs,superscriptaddress,floatfix,nofootinbib]{revtex4-2}

\usepackage{graphicx,graphics}
\usepackage{dcolumn}
\usepackage{amsmath,amssymb,amsfonts}
\usepackage{latexsym,verbatim}
\usepackage{bm}
\usepackage{breqn}

\makeatletter
\let\cat@comma@active\@empty
\makeatother

\usepackage[toc,page]{appendix}
\usepackage{color}
\usepackage{ulem}
\usepackage{tikz}
\usepackage[breaklinks=true,colorlinks,citecolor=blue,linkcolor=blue,urlcolor=blue]{hyperref}
\usepackage{tabularx}
\usepackage{array, multirow, bigdelim, makecell, booktabs} 
\graphicspath{fig}

\begin{document}

\title{Broken-Symmetry Ground States of the Heisenberg Model on the Pyrochlore Lattice}

\author{Nikita~Astrakhantsev}
\email[]{nikita.astrakhantsev@physik.uzh.ch}
\affiliation{Department of Physics, University of Zurich, Winterthurerstrasse 190, CH-8057 Zurich, Switzerland}

\author{Tom~Westerhout}
\email[]{tom.westerhout@ru.nl}
\affiliation{Institute for Molecules and Materials, Radboud University,
Heyendaalseweg 135, 6525AJ Nijmegen, The Netherlands}

\author{Apoorv~Tiwari}
\affiliation{Department of Physics, University of Zurich, Winterthurerstrasse 190, CH-8057 Zurich, Switzerland}
\affiliation{Condensed Matter Theory Group, Paul Scherrer Institute, CH-5232 Villigen PSI, Switzerland}

\author{Kenny~Choo}
\affiliation{Department of Physics, University of Zurich, Winterthurerstrasse 190, CH-8057 Zurich, Switzerland}

\author{Ao~Chen}
\affiliation{Institute for Theoretical Physics, ETH Zurich, CH-8093 Zurich, Switzerland}

\author{Mark~H.\,Fischer}
\affiliation{Department of Physics, University of Zurich, Winterthurerstrasse 190, CH-8057 Zurich, Switzerland}

\author{Giuseppe~Carleo}
\affiliation{Institute of Physics, École Polytechnique Fédérale de Lausanne (EPFL), CH-1015 Lausanne, Switzerland}

\author{Titus~Neupert}
\affiliation{Department of Physics, University of Zurich, Winterthurerstrasse 190, CH-8057 Zurich, Switzerland}

\begin{abstract}
The spin-1/2 Heisenberg model on the pyrochlore lattice is an iconic frustrated three-dimensional spin system with a rich phase diagram. 
Besides hosting several ordered phases, the model is debated to possess a spin-liquid ground state when only nearest-neighbor antiferromagnetic interactions are present.
Here, we contest this hypothesis with an extensive numerical investigation using both exact diagonalization and complementary variational techniques. Specifically, we employ a RVB-like many-variable Monte Carlo ansatz and convolutional neural network quantum states for (variational) calculations with up to $4\times 4^3$ and $4 \times 3^3$ spins, respectively. 
We demonstrate that these techniques yield consistent results, allowing for reliable extrapolations to the thermodynamic limit. {We consider the $(\lambda, j_2 / j_1)$ parameter space with $j_2,\,j_1$ being nearest and next-to-nearest neighbor interactions and $\lambda$ the XXZ interaction anisotropy.} Our main results are (1) the determination of the phase transition between the putative spin-liquid phase and the neighboring magnetically ordered phase and (2) a careful characterization of the ground state in terms of symmetry-breaking tendencies.
We find clear indications of {a dimer order with} spontaneously broken inversion and rotational symmetry, calling the scenario of a featureless quantum spin-liquid into question. Our work showcases how many-variable variational techniques can be used to make progress in answering challenging questions about three-dimensional frustrated quantum magnets. 
\end{abstract}
\maketitle

\section{Introduction} 
Featureless  ground states of interacting quantum spins with exotic properties and emergent excitations are highly sought after. Identifying such quantum spin liquids (QSL) is particularly challenging in three-dimensional (3D) systems due to fast scaling of the Hilbert space with linear system size. A key ingredient that favors liquid over ordered phases is geometrical frustration, with the pyrochlore lattice a prominent frustrated 3D lattice.
On the pyrochlore lattice, even the (classical) antiferromagnetic spin-1/2 Ising model has a ground-state manifold with an extensive degeneracy, governed by the ``two-in, two-out'' spin ice rule, instead of magnetic order~\cite{Singh_2012,PhysRevB.43.865,PhysRevLett.80.2929,PhysRevB.58.12049}. Upon inclusion of a small exchange term, excitations with fractionalized ``magnetic'' charges and gauge photons emerge, the hallmarks of a $U(1)$ QSL ~\cite{PhysRevX.1.021002,hermele2004pyrochlore,savary2012coulombic,PhysRevX.7.041057,Benton_2018,PhysRevLett.116.177203}.

At the same time, the nature of the ground state away from perturbative limits around the Ising point on the pyrochlore lattice is subject to long-standing debates. A prominent example is the ground state of the $SU(2)$ Heisenberg nearest-neighbor antiferromagnet: Perturbative or mean-field treatments suggest the ground state to be dimerized~\cite{doi:10.1063/1.348098,doi:10.1143/JPSJ.67.4022,PhysRevB.63.144432,doi:10.1143/JPSJ.70.640,PhysRevB.65.024415,Lee_2002,Berg_2003,fouet2003planar,Moessner_2006,Normand_2014} or to possess chiral magnetic order~\cite{Kim_2008}. Alternatively, the aforementioned $U(1)$ QSL phase might continue from the small exchange interaction region to the $SU(2)$-symmetric point without a phase transition~\cite{iqbal2019quantum,PhysRevLett.80.2933,PhysRevB.61.1149,PhysRevLett.116.177203,PhysRevX.7.041057}.
Resolving this debate is a challenging theoretical problem with direct relevance to real compounds, such as rare earth molybdenum oxynitride pyrochlores $\mbox{R}_2\mbox{Mo}_2\mbox{O}_5\mbox{N}_2$, which are expected to admit an antiferromagnetic isotropic spin-1/2 Heisenberg model description~\cite{yang2010}. Remarkably, the recently synthesized $\mbox{Lu}_2\mbox{Mo}_2\mbox{O}_5\mbox{N}_2$ experimentally shows neither sign of magnetic order nor spin freezing~\cite{PhysRevMaterials.1.071201,PhysRevLett.113.117201}.

A QSL phase, if present, is expected to be close in parameter space to several symmetry-breaking ordered phases~\cite{PhysRevX.7.041049,Tsuneishi_2007,NAKAMURA20071297,PhysRevB.78.144418,PhysRevB.84.144432}. These ordered phases can be induced by including next-to-nearest neighbor couplings~\cite{PhysRevX.7.041049,iqbal2019quantum} or large transverse exchange interactions, with the latter stabilizing spin-nematic order~\cite{PhysRevX.7.041057,Benton_2018,PhysRevLett.80.2933,PhysRevB.61.1149}. This suggests a close competition between nearly degenerate QSL and ordered phases.

As for any 3D frustrated quantum magnet, reliable numerical methods, crucial to resolving such a competition of nearly degenerate states in the spin-1/2 Heisenberg pyrochlore model, are scarce. Perturbative approaches lead to disagreeing conclusions~\cite{doi:10.1063/1.348098,doi:10.1143/JPSJ.67.4022,PhysRevB.63.144432,doi:10.1143/JPSJ.70.640,PhysRevB.65.024415,Lee_2002,Berg_2003,fouet2003planar,Moessner_2006,Normand_2014}, while fermionic mean-field approaches~\cite{PhysRevB.78.180410} and the functional renormalization group (FRG) method~\cite{iqbal2019quantum}, lack control parameters to improve results systematically. Unbiased methods are either limited to small clusters, like exact diagonalization (ED)~\cite{PhysRevLett.80.2933,PhysRevB.97.144407}, only reach temperatures of order $T \sim J / 6$, like diagrammatic Monte Carlo~\cite{PhysRevLett.116.177203}, or are limited to non-frustrated exchange terms, like quantum Monte Carlo methods~\cite{PhysRevB.43.5950,PhysRev.138.A442}. Finally, density matrix renormalization group (DMRG) calculations are plagued by the cut dimensionality in 3D systems~\cite{RevModPhys.77.259}, even though significant advances have been demonstrated~\cite{hagymasi2020possible}. Further numerical approaches are thus needed to settle the ground-state question in the spin-1/2 Heisenberg model on the pyrochlore lattice.

In this work, we assess whether the spin-1/2 pyrochlore quantum antiferromagnet hosts a QSL or a symmetry-broken ground state~\cite{hagymasi2020possible} in the vicinity of the $SU(2)$--symmetric point and further identify the adjacent phases. For this we use state-of-the-art variational Monte Carlo (VMC) methods. We use two \textit{complementary} and highly flexible ansatz wave functions, thus controlling the inherent parametrical bias. On the one hand, we use the  many-variable variational Monte Carlo (mVMC) method inspired by the resonating-valence-bond (RVB) wave function~\cite{misawa2019mvmc}, and recently adopted for two-dimensional frustrated models~\cite{nomura2020dirac}. The second ansatz uses neural network quantum states (NQS)~\cite{carleo2017solving}, which have proven to be a powerful addition to the numerical toolbox for many-body quantum systems~\cite{PhysRevB.100.125124,nomura2020dirac}. Finally, ED calculations (we employ a novel package \texttt{SpinED}~\cite{westerhout2021latticesymmetries}) are used to benchmark both approaches and guide our analysis.

\begin{figure}[t!]
    \centering
    \begin{tikzpicture}
        \node[inner sep=0pt] at (0.0, 0) {\includegraphics[width=0.50\textwidth]{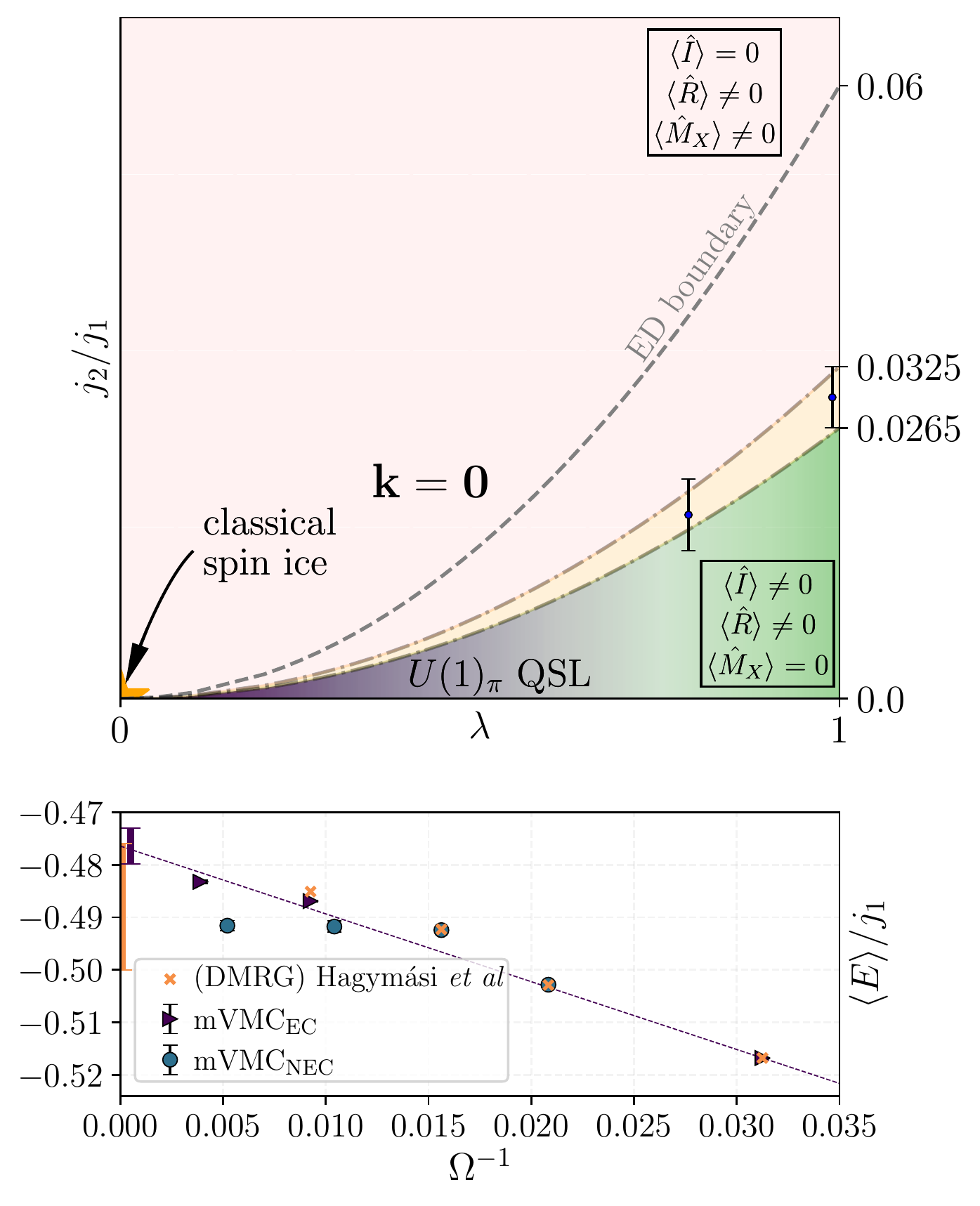}};
        \node[inner sep=0pt] at (-1.1, 3.3)     {\includegraphics[width=0.25\textwidth]{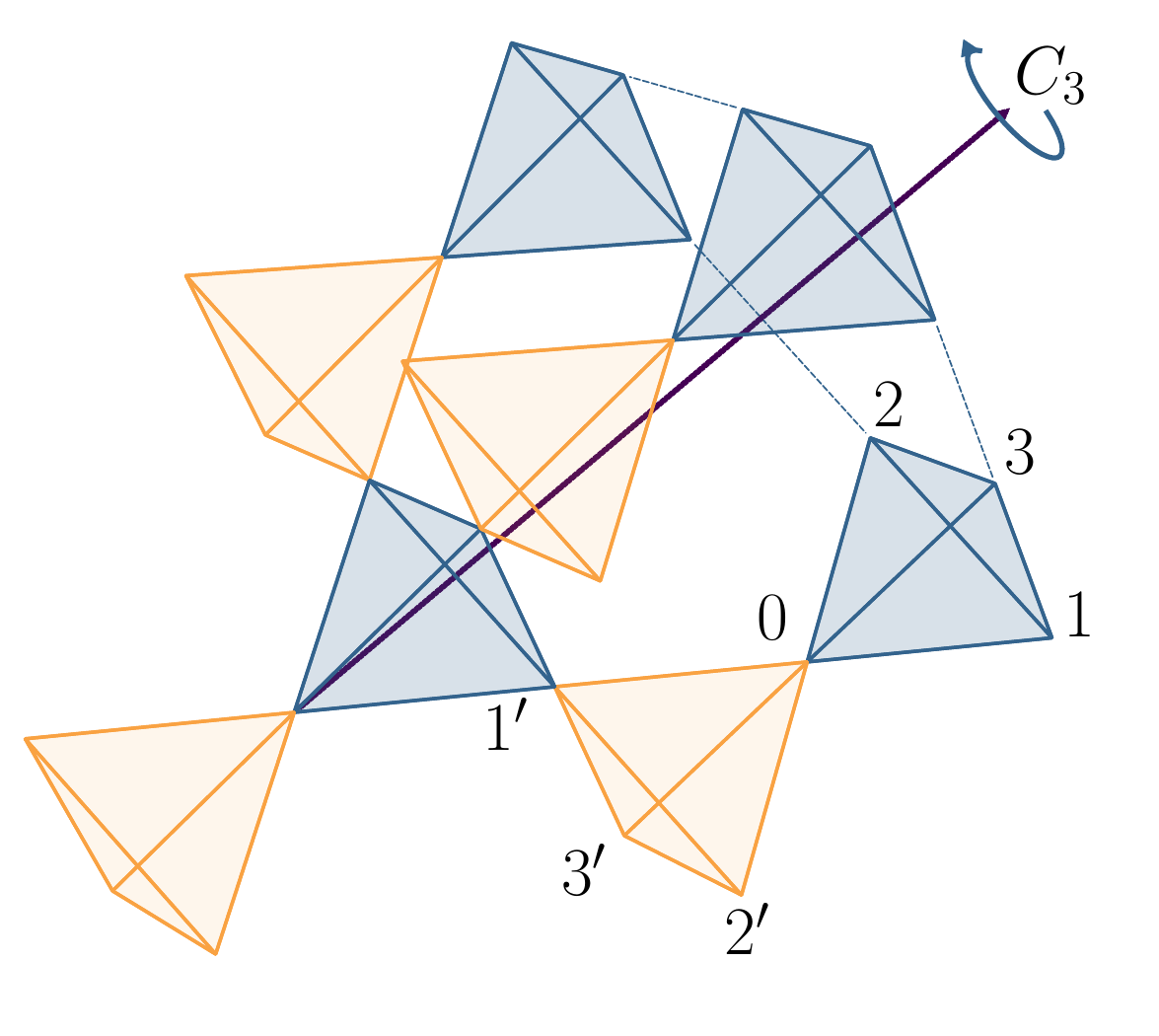}};
        \node at (-0.3, 5.55)  {(a) Phase diagram};
        \node at (-0.3, -1.65)  {(b) Energy extrapolation at $j_2 / j_1 = 0$, $\lambda = 1$};
    \end{tikzpicture}
    \caption{({a}) Pyrochlore Heisenberg $\lambda$ - $j_2 / j_1$ phase diagram.
    The non-magnetic phase (green) in the vicinity of $(1,\,0)$ is separated by a phase transition (yellow) from the magnetically ordered $\mathbf k = \mathbf 0$ phase (red). The $U(1)_{\pi}$ QSL phase is shown in grey. Further indicated are the symmetry-breaking properties of the respective phases with respect to inversion ($\hat{I}$) and rotation ($\hat{R}$) symmetry.
    The ED phase boundary was obtained for a $4\times 2^3$ system. Blue data points with errorbars show our infinite-volume phase-boundary extrapolations at $\lambda = 0.8$ and $1.0$. {Note that our data does not allow us to reliably draw conclusion about the shape of the phase boundary in the region $\lambda \ll 1$. The shape of the yellow region, shown in Fig.~\ref{fig:intro}~(a) at $\lambda \lesssim 0.8$ is a guide to the eye consistent with perturbative results for small $\lambda$.}
    The inset in the upper left shows the pyrochlore structure with sublattice indices. 
    ({b}) The best variational energies obtained within this work using mVMC on equilateral (EC) and non-equilateral (NEC) clusters at $\lambda = 1$, $j_2 / j_1$ = 0. The dashed line represents the infinite-volume extrapolation. For comparison, the figure also shows the DMRG result of Ref.~\cite{hagymasi2020possible} without bond-dimension extrapolation. The errorbars at $\Omega^{-1} = 0$ represent the infinite-volume extrapolations obtained within this work and DMRG bond-dimension extrapolation of Ref.~\cite{hagymasi2020possible}, see main text for details.
    }
    \label{fig:intro}
\end{figure}

Our methodology allows us to obtain reliable wave functions on clusters as large as $256 = 4 \times 4^3$ sites and perform extrapolations to the thermodynamic limit, which is notoriously hard for a 3D frustrated system. Figure~\ref{fig:intro} summarizes our main results in the form of (a) the phase diagram and (b) our energy extrapolation to the thermodynamic limit. 
In the putative QSL phase, we observe clear signatures suggestive of spontaneous breaking of both inversion and rotation symmetry. 
Furthermore, we extrapolate the phase transition of this non-magnetic symmetry broken phase with the adjacent magnetic phase upon addition of next-nearest neighbor coupling $j_2$. We find the phase transition at 7 times smaller $j_2/j_1$ as compared to previous FRG results~\cite{iqbal2019quantum}, which expands the magnetically ordered phase significantly. 

The rest of this paper is organized as follows: In Sec.~\ref{sec:methodology}, we introduce the model and our methodology. In Sec.~\ref{sec:results}, we discuss our results, starting with a validation of the method by combining results from ED with the two variational approaches and a characterization of the magnetically ordered phase. This is followed by a thorough analysis of the symmetry-breaking signatures in the non-magnetic phase. Finally, Sec.~\ref{sec:discussion} concludes with a discussion.

\section{Model and Methodology}
\label{sec:methodology}

\subsection{Model and system geometry}
\label{subsec:geometry}
We consider the interacting spin Hamiltonian 
\begin{eqnarray}
    \hat H(\lambda, j_1, j_2) &=& j_1 \sum\limits_{\langle i, j\rangle} \hat h_{\lambda}(i, j) + j_2 \sum\limits_{\langle \langle i, j\rangle \rangle} \hat h_{\lambda}(i, j),\\
        \hat h_{\lambda}(i, j) &=& \frac{\lambda}{2} \left( \hat S^+_i \hat S^-_j + \hat S^+_j \hat S^-_i\right) + \hat S^z_i \hat S^z_j,
    \label{eq:hamiltonian}
\end{eqnarray}
where $\langle \ldots \rangle$ denotes summation over nearest neighbors, $\langle \langle \ldots \rangle \rangle$ over next-to-nearest neighbors on the pyrochlore lattice {and the local quantization axis of spin points towards the center of a tetrahedron.} In this work, we are mainly interested in determining the $(\lambda,\,j_2 / j_1)$ phase diagram. Two limiting cases of the phase diagram are readily identified. On the one hand, the point $(\lambda = 0,\,j_2/j_1 = 0)$ corresponds to the classical Ising model which hosts the spin-ice ground state manifold. On the other hand, $\lambda = 1$ corresponds to the $SU(2)$ symmetric Heisenberg model. Fragments of this phase diagram, which is schematically depicted in Fig.~\ref{fig:intro}~(a), were previously studied. The classical Ising model orders magnetically upon inclusion of a small $j_2 / j_1$ term~\cite{iqbal2019quantum,Tsuneishi_2007,NAKAMURA20071297,PhysRevB.78.144418,PhysRevB.84.144432}. The so-called $\mathbf{k}= \mathbf 0$ magnetic order does not break translation symmetry, while the magnetic moments within each of the four sublattices are ordered ferromagnetically, such that the total magnetization within each tetrahedron vanishes. On the $j_2/j_1 = 0$ axis, perturbation theory in $\lambda > 0$ reveals the $U(1)_{\pi}$ QSL phase~\cite{hermele2004pyrochlore,Savary_2016QSI,Chen_2017}.

In order to better clarify the symmetries of the problem, we start by discussing the geometry of the pyrochlore lattice. The inset in Fig.~\ref{fig:intro}~(a) shows the pyrochlore crystal structure. The pyrochlore lattice $\Lambda$ is the union of four sublattices $\Lambda_\alpha$, $\alpha=0,\,1,\,2,\,3$, which in Cartesian coordinates are spanned by the unit vectors $\bm e_1 = (1,\,1,\,0)^T,$ $\bm e_2 = (0,\,1,\,1)^T,$ $\bm e_3 = (1,\,0,\,1)^T$,
\begin{equation}
    \Lambda_\alpha=\left\{\left.
    \sum_{\beta=1}^3 n_\beta \bm{e}_\beta+\bm{e}_\alpha/2 \right|n_\beta=0,1,\cdots, L_\beta-1
    \right\}.
\end{equation}
Here, $\bm{e}_0=(0,\,0,\,0)^T$ and $L_1$,\,$L_2$,\,$L_3$ are the number of unit cells in the three respective directions that span the lattice. In each unit cell, the resulting four sites form a tetrahedron. Note that each site of this lattice is shared between exactly two tetrahedra. In what follows, we refer to ``up-tetrahedra'' and ``down-tetrahedra'' as indicated in blue and orange, respectively in Fig.~\ref{fig:intro}~(a). The total number of sites is denoted by $\Omega$ throughout the manuscript.

We consider pyrochlore clusters with periodic boundary conditions. Their number of bonds is 12 times the number of unit cells. We associate with one unit cell the four lattice sites labelled 0,\,1,\,2,\,3 in Fig.~\ref{fig:intro}~(a) as well as the 12 bonds comprising the up- and down-tetrahedra that have their corners labelled.
The space group of the pyrochlore lattice is $Fd\bar{3}m$ with a point group isomophic to $O_h$. In the following, to study symmetry breaking tendencies, we will mainly focus on equilateral pyrochlore clusters  ($L_1 = L_2 = L_3$), which have a point group $D_{3d}$. It is generated by $C_3$ rotations around the ``easy-axis'' $(1,\,1,\,1)^T$, which cyclically exchange the sites ($1 \to 2 \to 3 \to 1)$ [see Fig.~\ref{fig:intro}~(a)], inversion symmetry $\;\mathbf r \to - \mathbf r$, which also exchanges down- and up-tetrahedra and a mirror symmetry, which reflects the cluster with respect to the plane passing through the bond 23 and the middle of the bond 01~\cite{Hwang_2020}. Importantly, such equilateral clusters avoid geometric bias for the symmetries studied in our calculations.

\subsection{Many-variable wave function}
The many-variable variational Monte Carlo (mVMC) method has proven successful in studies of strongly correlated phases~\cite{PhysRevB.90.115137,casula2004correlated}, including the QSL and valence bond solid (VBS) in the $j_1$-$j_2$ Heisenberg model on the square lattice~\cite{doi:10.7566/JPSJ.84.024720,nomura2020dirac}. Here, we use the highly-optimized realization of mVMC from Ref.~\cite{misawa2019mvmc,doi:10.1143/JPSJ.77.114701}. At the heart of the mVMC method is a mapping of spin operators to fermionic bilinears 
\begin{equation}
    \hat S^a_i \to \frac{1}{2} \sum_{\alpha,\beta=\uparrow, \downarrow}\hat  c^{\dagger}_{i,\alpha} \sigma^a_{\alpha \beta} \hat  c_{i,\beta},
\end{equation}
where $i$ labels a lattice site, $a=x,\,y,\,z$, and $\sigma^x$, $\sigma^y$, $\sigma^z$ are the three Pauli matrices. 

The RVB-like pairing state has the form
\begin{equation}
    |\phi_{\mbox{\footnotesize pair}} \rangle = \mathcal{P}^{\infty}_{\mbox{\footnotesize G}} \exp \left(\sum\limits_{i, j} f_{i,j} \hat c^{\dagger}_{i, \uparrow} \hat c^{\dagger}_{j \downarrow} \right) |0 \rangle,
\end{equation}
where single occupation is ensured by the $\mathcal{P}^{\infty}_{\mbox{\footnotesize G}}$ single-occupation Gutzwiller projector. Note that the $\mathcal{P}^{\infty}_{\mbox{\footnotesize G}}$--projected fermionic Hilbert space can be mapped to the original Hilbert space of spin operators. The wave-function value $\langle \boldsymbol{\sigma} |\phi_{\mbox{\footnotesize pair}} \rangle $ of a specific spin configuration $|\boldsymbol{\sigma}\rangle$ is evaluated using the Slater determinant of the matrix with elements $f_{i,j}$. Here, $\boldsymbol{\sigma}$ represents a string of $\pm 1$, which, for each lattice site,  stands for the respective spin eigenstate in the $S^z$ basis. The parameters $f_{i,j}$ are optimised using the stochastic reconfiguration optimization technique~\cite{sorella_green_1998}, which can also be seen as a way of performing stochastic imaginary-time dynamics in the variational manifold~\cite{becca_quantum_2017,carleo2017solving}.

\subsection{Neural Network Quantum States}
To support the physical conclusions obtained within one parametrical ansatz, we complement our study with a completely different variational wave function, namely an NQS. Recently, NQS have been applied to obtain a successful parametrization of the Heisenberg-model wave function~\cite{carleo2017solving}, which ignited the development of NQS as a broadly applicable variational method~\cite{westerhout2019neural,burau2020unitary,verdel2020variational,nomura2020dirac,nomura2020helping,carrasquilla2017machine,torlai2018neural,ferrari2020variational,ferrari2019neural,Szab__2020,PhysRevLett.124.020503,PhysRevResearch.2.012039}. Another example of a recent application of NQS are 2D frustrated magnets~\cite{PhysRevB.100.125124,westerhout2019neural}. In this work, we employ the comprehensive \texttt{NetKet} NQS implementation~\cite{netket:2019}.

The general idea of the NQS method is to use the spin configuration $\boldsymbol{\sigma}$ as input for a neural network $\Psi$, interpreting the result $\Psi(\boldsymbol{\sigma})$ as the (not normalized) wave function component, corresponding to the basis vector $\boldsymbol{\sigma}.$ To cope with the large system size and possible overfitting~\cite{westerhout2019neural}, we employ the {\it convolutional} neural network architecture (CNN) with real parameters and an elaborate alternating training technique (see Appendix~\ref{sec:appendix_NQS} for details). This choice of architecture allows for better optimization and avoids certain instabilities~\cite{bukov2020learning}. We also benchmark it against the Restricted Boltzmann Machine (RBM) network, comprising one fully-connected dense layer, which is the classic NQS architecture~\cite{nomura2020helping,carleo2017solving}. The details of the NQS architectures and the variational parameter training can be found in Appendix~\ref{sec:appendix_NQS}.

\subsection{Symmetry-projected wave functions}
\label{subsec:spwf}
To obtain highly accurate variational wave functions and energies, we employ quantum-number projections, i.e., we impose the ansatz state to transform in a chosen irreducible representation of the symmetry group. 

Any point-group symmetry $\hat G$ is projected by applying it until the symmetry orbit is exhausted
\begin{equation}
 |\Psi_\xi \rangle = \hat P |\Psi \rangle =  \sum\limits_n \xi^n \hat G^n |\Psi \rangle,
    \label{eq:q_projection}
\end{equation}
where $\xi$ is the desired projection quantum number and $|\Psi_\xi \rangle$ the symmetrized state. 

Similarly, within mVMC, the projection onto the total spin $S$ is performed by superposing the $SU(2)$--rotated wave functions~\cite{doi:10.1143/JPSJ.77.114701}. {At $\lambda \neq 1$, when only $U_z(1)$ spin rotation symmetry is present, the symmetry is enforced by only working in the space of total-zero magnetization $\hat M_z |\psi \rangle = 0.$} In the NQS method, the spin-parity projector is applied:
\begin{equation}
 \Psi_{\pm} (\boldsymbol{\sigma}) = \Psi (\boldsymbol{\sigma}) \pm \Psi (-\boldsymbol{\sigma}),
    \label{eq:parity_projection}
\end{equation}
where $\Psi(\pm \boldsymbol{\sigma})$ is the wave function evaluated at spin configurations $\boldsymbol{\sigma}$ and the spin configuration flipped along the $z$-axis, $-\boldsymbol{\sigma}$. Such a projector selects wave functions of either even or odd total spin.

For the mVMC ansatz, momentum and point-group-symmetry projection can partially be performed by directly constraining the variational parameters $f_{i,\,j}$. Otherwise, taking all $\Omega^2$ parameters $f_{i,\,j}$ independent and symmetrizing the wave function at the end leads to a prohibitively large number of terms in the projector. In addition to computational cost, this makes the optimization procedure prone to false minima convergence. 
As a compromise, for clusters larger than $48$ sites, we impose translational symmetry on the variational parameters $f_{i,\,j}$ and project the other symmetries using Eq.\,\eqref{eq:q_projection}. Similarly, within the NQS method, the CNN architecture automatically imposes translational symmetry on the network parameters, while other projections are done using Eq.\,\eqref{eq:q_projection}. To avoid the possible false minima convergence, in both methods we employ long Monte Carlo samples and try dozens of random initial approximations to select the best energy (see Appendix\,\ref{sec:appendix_NQS}). 

\subsection{Symmetry-breaking susceptibilities}
One of the most direct characterizations of an ordered phase is through its symmetry breaking pattern. Consider an operator $\hat{\mathcal{O}} = \Omega^{-1} \sum\limits_i \hat{o}_i$ with $\hat{o}_i$ acting locally, which measures symmetry breaking. In other words, the operator transforms non-trivially under the actions of the symmetry group. 
In a finite volume, data indicates that the pyrochlore ground state belongs to a trivial representation of all symmetries (point-group and spin), which forbids direct observation of the $\langle \hat{\mathcal{O}} \rangle \neq 0$ condensate. Instead, one measures the (equal-time) susceptibility $\chi_{\hat{\mathcal{O}}} = \langle\hat{\mathcal{O}}^{\dagger} \hat{\mathcal{O}} \rangle$, which vanishes or survives in the thermodynamic limit if the phase has a symmetric or symmetry-broken ground state, respectively. Note that a non-vanishing susceptibility can only arise due to the establishment of long-range order.

To probe the $SU(2)$--symmetry breaking through long-range magnetic order, we calculate $\chi_{\hat{M}_{\mathbf k}}$, introducing the $\mathbf k$--dependent operator 
\begin{equation}
    \hat M_{\mathbf k} = \frac{1}{\Omega} \sum\limits_{i} \hat S_i^z e^{i \mathbf k \mathbf{r}_i},
\end{equation}
where $\mathbf{r}_i$ is the physical position of the site $i$ and $\mathbf{k}$ takes values in the extended Brillouin zone. Since the Hamiltonian is $SU(2)$--symmetric, we restrict ourselves to the $z$--spin component only.

To probe the point-group symmetry-breaking tendency in the absence of magnetic order, we construct dimer-type operators
\begin{align}
   \begin{split}
    & \hat{\mathcal{O}}(\xi, \omega)=  
    \frac{1}{3 \Omega} \sum\limits_{\langle i ,j\rangle} 
    q_{i,j}(\xi, \omega)\,
    \hat{\mathbf S}_{i} \cdot \hat{\mathbf S}_{j},
   \end{split}
   \label{eq:inversion_dimer}
\end{align}
where 
\begin{equation}
    \begin{split}
        q_{i,\,j}(\xi, \omega)= \begin{cases}
        \omega^{\ell-1} & \mathbf{r}_i-\bm{r}_j=+\mathbf{e}_\ell/2\\
        \xi\,\omega^{\ell-1} & \mathbf{r}_i-\bm{r}_j=-\mathbf{e}_\ell/2\\
        \omega^{\ell} & \mathbf{r}_i-\bm{r}_j=+\mathbf{e}_\ell/2 - \mathbf{e}_{\ell - 1}/2\\
        \xi \omega^{\ell} & \mathbf{r}_i-\bm{r}_j=-\mathbf{e}_\ell/2 + \mathbf{e}_{\ell - 1}/2\\
        \end{cases}
    \end{split}
    \label{eq:relative_phases}
\end{equation}
with $\ell=1,\,2,\,3$.
Here, $\xi \in \{\pm 1\}$ and $\omega \in \{1,\,\exp \left(\pm 2 \pi i / 3 \right)\}$ are the eigenvalues of inversion and rotation, respectively. 
Condensation of $\hat{\mathcal{O}}$ with non-trivial $\xi$ or $\omega$ signals symmetry breaking. Note that the $C_3$ rotation does not mix the bond groups $\{(0,\,i),\;1 \leqslant i \leqslant 3\}$ and $\{(j,\,i),\;1 \leqslant i,\;j \leqslant 3\}$, so the specific eigenvalue of $C_3$ does not fix the relative phase between these groups. The phases given in\,\eqref{eq:relative_phases} are equal on opposite bonds within a tetrahedron, which is suggested by the dimer-dimer correlation pattern obtained within ED.
Practically, such choice improves the signal-to-noise ratio in Monte-Carlo measurements of the susceptibility.

Similarly, in~\eqref{eq:inversion_dimer}, we define $\hat{\mathcal{O}}(\zeta)$ with
\begin{equation}
    \begin{split}
        q_{i,\,j}(\zeta)= \begin{cases}
        (1 + \zeta) / 2 & \mathbf{r}_i-\bm{r}_j=\pm (\mathbf{e}_2 - \mathbf{e}_3)\\
        (1 + \zeta) / 2 & \mathbf{r}_i-\bm{r}_j=\pm (\mathbf{e}_0 - \mathbf{e}_1)\\
        1 & \mathbf{r}_i-\bm{r}_j=\pm (\mathbf{e}_0 - \mathbf{e}_\ell)\\
        \zeta & \mathbf{r}_i-\bm{r}_j=\pm (\mathbf{e}_1 - \mathbf{e}_\ell),\\
        \end{cases}
    \end{split}
    \label{eq:relative_phases_sigma}
\end{equation}
where $\ell=2,\,3$, to probe spontaneous breaking of the mirror symmetry. Here $\zeta \in \{\pm 1\}$ is the mirror eigenvalue.

\section{Results}
\label{sec:results}
We present our results in the following order: First, we demonstrate that the variational energies we obtain compare favorably to the previous studies and then show finite size extrapolations based on computations with system sizes beyond those available in the literature. We show that our variational wave functions correctly capture the magnetic order in the $\mathbf k = \mathbf 0$ phase and the absence of magnetic order at the nearest-neighbor Heisenberg point. Second, we present our finite size extrapolation of the transition point between the two phases along the $j_2/j_1$ axis for $\lambda=1$. Third, we discuss the results of order-parameter and susceptibility calculations, elucidating the symmetry-breaking characteristics of both phases.

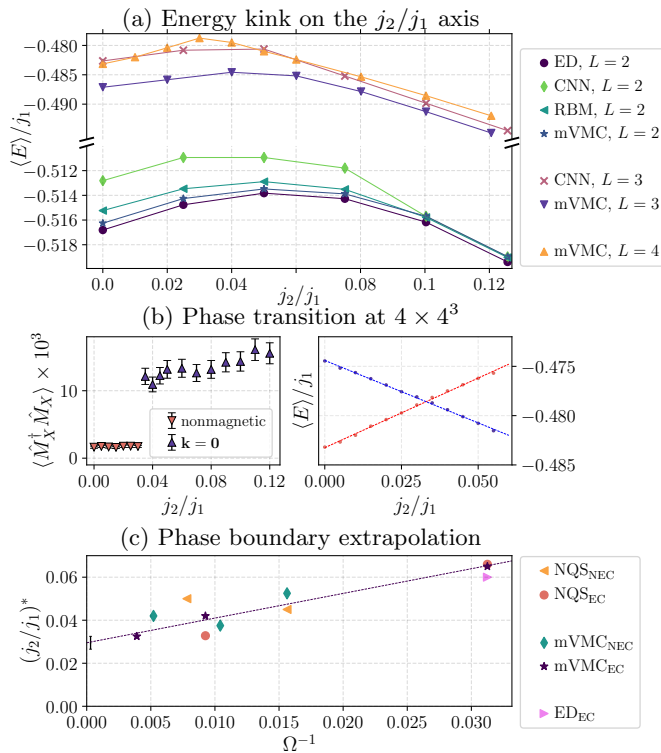
\begin{figure}[t!]
    \centering
    \begin{tikzpicture}
        \node[inner sep=0pt] at (0, 0)   {\includegraphics[width=0.50\textwidth]{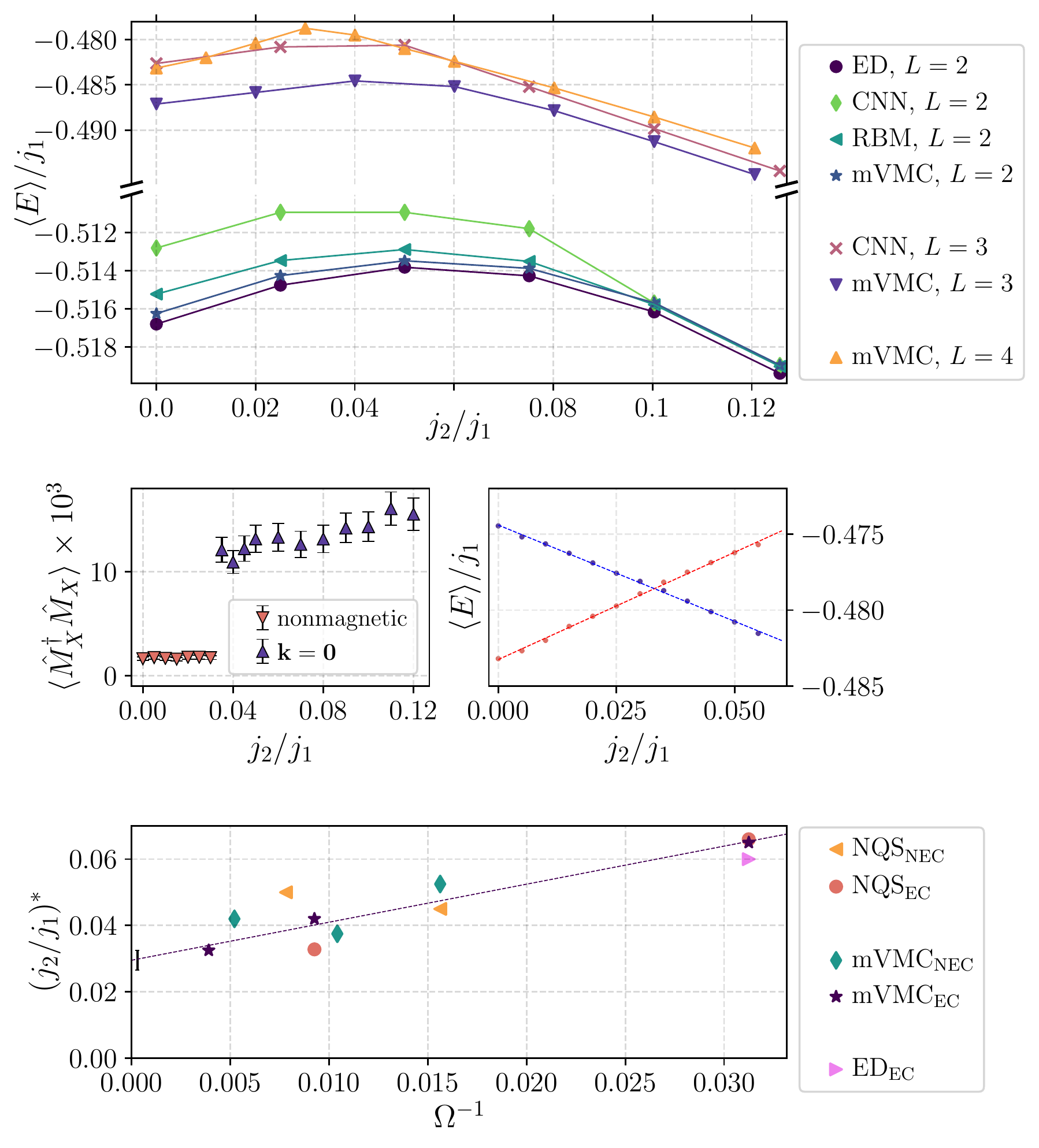}};
        \node at (-0.5, 4.95)  {(a) Energy kink on the $j_2 / j_1$ axis};
        \node at (-0.5, 1.00)  {(b) Phase transition at $4 \times 4^3$};
        \node at (-0.5, -1.95)  {(c) Phase boundary extrapolation};
    \end{tikzpicture}
    \caption{Phase transition at the $SU(2)$-symmetric point and its extrapolation to the thermodynamic limit. (a)~Ground state energy on $4 \times L^3$ clusters obtained within the various methods. (b)~Left panel: magnetic susceptibility at $\mathbf k = X$ as the function of $j_2 / j_1$ within the hysteresis optimization performed at the $4 \times 4^4$ cluster within mVMC. Right panel: energy level crossing obtained using the hysteresis optimization. (c)~Position of the phase boundary $(j_2 / j_1)^*$ for equilateral clusters (EC) or non-equilateral clusters (NEC) as the function of inverse cluster size. The ED point corresponds to the position of the energy maximum. The dashed line is the linear fit over $\mbox{mVMC}_{\mbox{\tiny EC}}$ points. The errorbar at $\Omega^{-1} = 0$ is conservatively estimated as difference between extrapolation and the measured transition point at the $4 \times 4^4$ cluster within mVMC.}
    \label{fig:corr_extrapolations}
\end{figure}

\subsection{Accuracy of wave functions}
\label{subsec:results_quality}
The accuracy of the wave functions we obtained can be compared with recent DMRG data on clusters up to $4 \times 3^3$~\cite{hagymasi2020possible} at the most frustrated $\lambda = 1$, $j_2 / j_1 = 0$ point. In Fig.~\ref{fig:intro}~(b) we show the best energy values obtained within DMRG and within our work. The mVMC errorbars arise from statistical uncertainty and the errorbar at $\Omega^{-1} = 0$ is estimated as $1/2$ absolute difference between the value obtained on the largest $4 \times 4^3$ cluster and the extrapolation result~\cite{hagymasi2020possible}. The energies are listed in Table~\ref{tab:energies_spingaps} of Appendix~\ref{sec:appendix_table}. On all clusters for which DMRG data is available, our variational energies agree with or are lower than the ones obtained by DMRG~\footnote{As the DMRG results, here we show the best actually obtained energy and not the bond-dimension extrapolation. We note that the infinite bond-dimension extrapolations, performed in~\cite{hagymasi2020possible}, slightly outperform our results by $2 \sigma$ and $\sigma$ on $4 \times 2^2 \times 4$ and $4 \times 3^4$ clusters, respectively.}. Our result $-0.477(3)$ for the infinite volume extrapolation is in agreement with the extrapolation from the DMRG calculations $-0.488(12)$ and improves upon the previous Variational Gutzwiller-projected mean field result~\cite{PhysRevB.79.144432}. This independent benchmark with DMRG confirms the ability of the variational Monte Carlo method to express the frustrated ground state even in the large volume.

Frustrated systems typically host many competing phases in a small energy window~\cite{nomura2020dirac}, thus a favorable comparison of energies is not always a guarantee of accuracy for a given variational ground state. However, we have further verified that physical observables obtained within the two variational parametrizations --- mVMC and NQS --- are also in striking agreement. To this end, we have studied the physical observables along the phase transition between the putative QSL phase at $j_2 / j_1 = 0$ and magnetically ordered $\mathbf k = \mathbf 0$ phase at large $j_2 / j_1$ (details about the $\mathbf k = \mathbf 0$  phase can be found in Appendix\,\ref{sec:appendix_k0}). 

\begin{figure*}[t!]
    \centering
    \begin{tikzpicture}
        \node[inner sep=0pt] at (-0.4, 0)   {\includegraphics[width=0.75\textwidth]{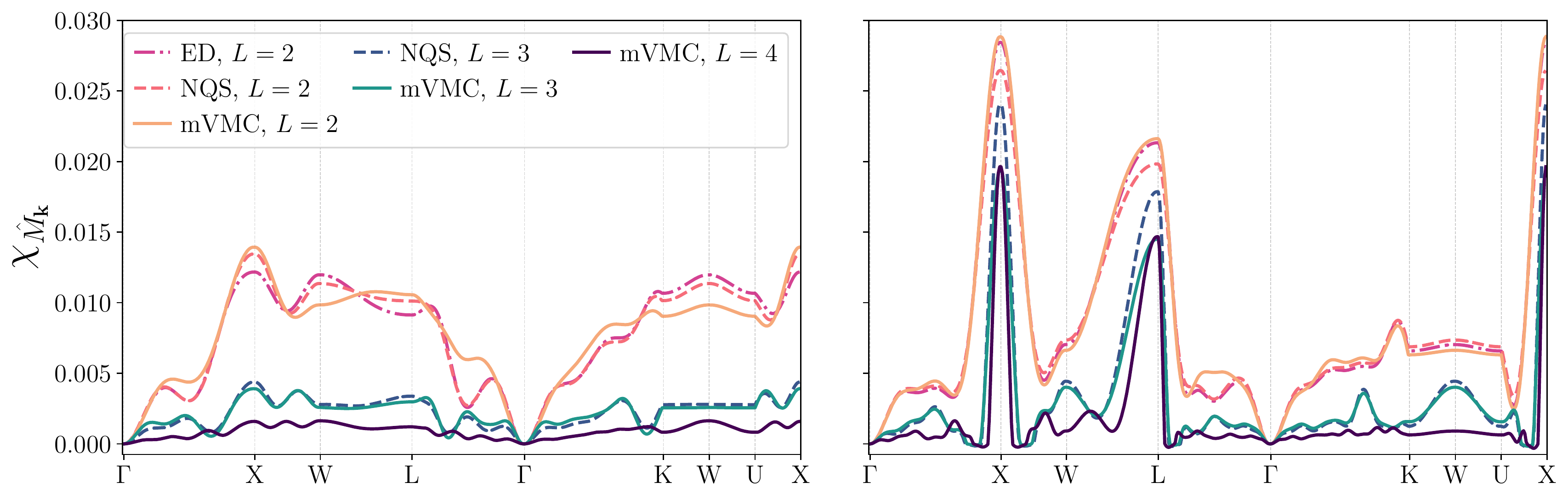}};
        \node[inner sep=0pt] at (8.9, 0)     {\includegraphics[width=0.27\textwidth]{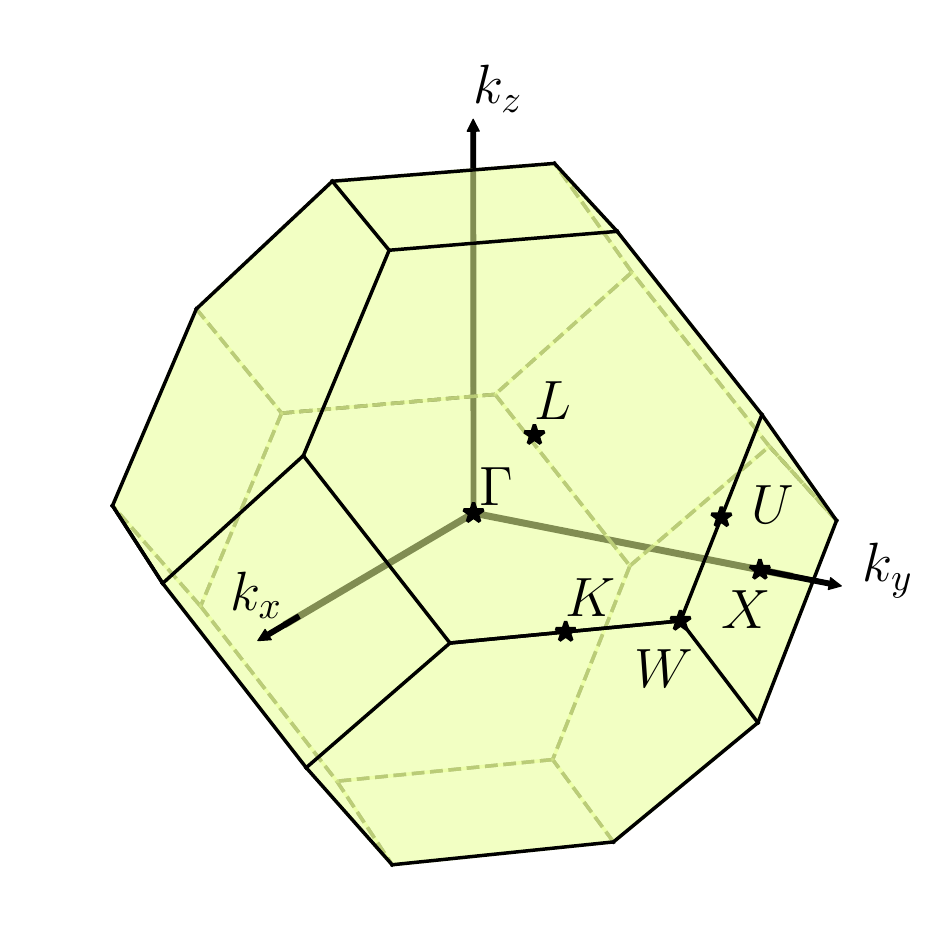}};
        \node at (-3.4, 2.25)  {(a) Susceptibility profile at $j_2 / j_1 = 0$};
        \node at (3.3, 2.25)  {(b) Susceptibility profile at $j_2 / j_1 = 0.25$};
        \node at (8.9, 2.25)  {(c) 3D extended Brillouin zone};
    \end{tikzpicture}
    \caption{Spin-spin susceptibility $\chi_{\hat{M}_{\mathbf k}}$ evaluated for the $SU(2)$--symmetric model { ($\lambda=1$)} for (a) the frustrated phase with $j_2 / j_1 = 0$ and (b) the $\textbf{k}= \mathbf 0$ magnetic phase with $j_2 / j_1 = 0.25$. The legend shows the methods and the size of the equilateral clusters used.\;(c) Extended Brillouin zone of the pyrochlore lattice with the high-symmetry points labeled.}
    \label{fig:corrs_hsl}
\end{figure*}
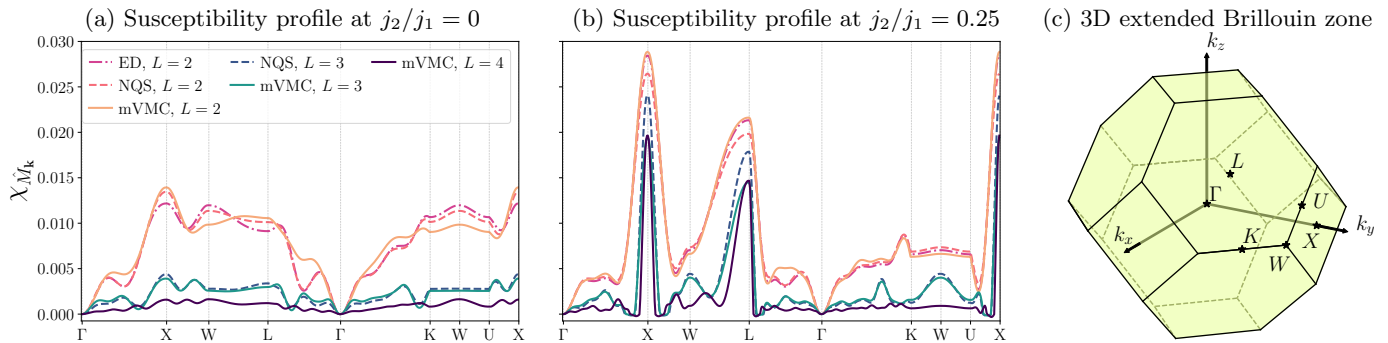

The ground-state energy behavior along the phase transition qualitatively agrees across all our approaches. Figure~\ref{fig:corr_extrapolations}~(a) shows the ground state energies at $\lambda = 1$ as a function of $j_2 / j_1$ obtained within ED and both variational methods on the equilateral clusters. The energy maximum signals the phase transition between the frustrated and the magnetic $\mathbf k = \mathbf 0$ phases. Note that a maximum is present for all approaches and clusters, but forms a more pronounced kink with larger cluster size, indicating a phase transition of first order. 

The NQS CNN results look slightly off in the frustrated phase. However, at $L = 2$ the CNN energy falls between first and second excited states\footnote{The relative NQS CNN energy error equals $\delta E / E_{\mbox{\footnotesize ED}} = 0.65$\,\% at $j_2 / j_1 = 0$, which outperforms the state-of-the-art result obtained for two-dimensional $j_2 / j_1$ square lattice Heisenberg model using CNN~\cite{PhysRevB.100.125124} and shows that this may be close to the limit given by the expressivity of the architecture.}, while overlap with the ground state is $|\langle \psi_{\mbox{\tiny ED}} | \psi_{\mbox{\tiny NQS}} \rangle| = 0.86$, which suggests that observables computed with $| \psi_{\mbox{\tiny NQS}} \rangle$ will be dominated by the system ground state.

Further support that the wave functions are consistent and physically correct can be obtained by comparison of magnetic susceptibilities towards $SU(2)$ spin-rotation symmetry breaking. In Fig.~\ref{fig:corrs_hsl}~(a--b) we show the magnetic susceptibility $\chi_{\hat{M}_{\mathbf k}}$ plotted along the high-symmetry line in the 3D extended Brillouin zone. Both variational methods not only agree with ED and among themselves, but also clearly distinguish between the non-magnetic phase in Fig.~\ref{fig:corrs_hsl}~(a) and the magnetically ordered $\mathbf k = \mathbf 0$ phase in Fig.~\ref{fig:corrs_hsl}~(b) in accordance  with the features described in Appendix\,\ref{sec:appendix_k0}, such as peaks at $\mathbf k = X,\,L$ and ratio between their heights.

\subsection{Magnetic phase boundary}
\label{subsec:results_boundary}
Based on these consistent results, we are able to  determine the shape of the phase boundary in the $(j_2 / j_1,\,\lambda)$ plane and extrapolate its location to the thermodynamic limit. 
As the position of the phase boundary, we take the maximum $j_2 / j_1$ such that the susceptibility on the whole $XW$ line is higher than the peak's half maximum (see Fig.~\ref{fig:corrs_hsl})\footnote{This is equivalent to the requirement that the full width at half maximum (FWHM) of the peak at $\mathbf k = X$ in the magnetic susceptibility is equal to the length of the $XW$ segment in the extended Brillouin zone.}. Applying this criterion to the ED data for $L = 2$ yields the phase diagram shown in Fig.~\ref{fig:intro}~(a). Within ED, the phase transition is located at $j_2 / j_1 \sim 0.06$ on the $SU(2)$-symmetric axis. Further, in the interval $0.09 \leqslant j_2 / j_1 \leqslant 0.12$ there occurs a level crossing of excitations belonging to $S = 1$ and $S = 0$ spin sectors, indicating a transition to a magnetic phase~\cite{nomura2020dirac,PhysRevLett.121.107202,PhysRevB.102.014417}. With decreasing $\lambda$, the non-magnetic phase width decreases within ED and vanishes at the classical spin ice point, meaning that even an infinitesimal positive $j_2 / j_1$ term breaks the spin-ice degeneracy and establishes magnetic order. The ED phase boundary is well described by a $(j_2 / j_1)^* = 0.06\lambda^2$ fit, shown as dashed line in Fig.~\ref{fig:intro}~(a). The FWHM-analysis details, the raw ED phase diagram, and level spectroscopy can be found in Appendix~\ref{sec:appendix_ED}.

A complementary, direct signature of the phase transition comes from the non-monotonous behavior (``kink'') of the energy as a function of $j_2 / j_1$, shown in Fig.~\ref{fig:corr_extrapolations}~(a).  
In the $L = 2$ ED-accessible case, the energy maximum is located at $j_2 / j_1 \sim 0.06$, which coincides with the FWHM analysis. Larger clusters with $L = 3$ and $L=4$ show similar behavior.  
To systematically obtain the kink position at large clusters, we perform a procedure comprising (1)~obtaining the best wave function at metastable points belonging surely to the frustrated ($j_2 / j_1 = 0.0$) or the ordered ($j_2 / j_1 = 0.2$) phases, (2)~adiabatically varying $j_2 / j_1$, adjusting the wave function, and lastly, (3)~locating the energy level crossing as an indication for the phase transition. An example of this \textit{hysteresis optimization} is shown in the right panel of Fig.~\ref{fig:corr_extrapolations}~(b). As the right panel shows, the energy level crossing is accompanied by an abrupt change in the ground state magnetic susceptibility. We find that the hysteresis optimization also significantly improves variational energies at the intermediate $j_2 / j_1$ interactions. For more details, see Appendix\,\ref{sec:appendix_hyst}.

Using the hysteresis optimization, we perform the phase boundary infinite-volume extrapolation at $\lambda = 1$ and $\lambda = 0.8$. We consider the geometries $4 \times 2^3$, $4 \times 2^2 \times 4$, $4 \times 2^2 \times 6$, $4 \times 2 \times 4^2$, $4 \times 2 \times 4 \times 6$, $4 \times 4^3$ and $4 \times 3^3$ with 32, 64, 96, 128, 196, 256 and 108 sites, respectively, but perform the extrapolation using equilateral clusters of the shape $4 \times L^3$ only. The transition points for all methods and geometries are shown in Fig.~\ref{fig:corr_extrapolations}~(c). Strikingly, the extrapolated range of the nonmagnetic phase at $\lambda = 1$ shrinks further as compared to the $4 \times 2^3$ system and the resulting critical value $(j_2 / j_1)^* = 0.0295(30)$ is an order of magnitude smaller than the result obtained from FRG, $(j_2 / j_1)^* = 0.22(3)$~\cite{iqbal2019quantum}. Note also that the phase-transition locations at the $4 \times 3^3$ cluster roughy agree within the two variational approaches, suggesting that the conclusions are not subject to strong variational bias. We use the difference between the mVMC data infinite volume extrapolation and the result obtained on the largest $4 \times 4^3$ cluster as the estimate of extrapolation uncertainty, which is shown as an errorbar in Fig.~\ref{fig:intro}~(a) and Fig.~\ref{fig:corr_extrapolations}~(c) at $\Omega^{-1} = 0$. Similarly, the $\lambda = 0.8$ point extrapolates to $(j_2 / j_1)^* = 0.0180(35)$. The extrapolation results at $\lambda = 0.8$ and $\lambda = 1$ are shown with error bars in Fig.~\ref{fig:intro}~(a). 
\begin{figure}[t!]
    \centering
    \begin{tikzpicture}
        \node[inner sep=0pt] at (-4.1, 0)    {\includegraphics[width=0.5\textwidth]{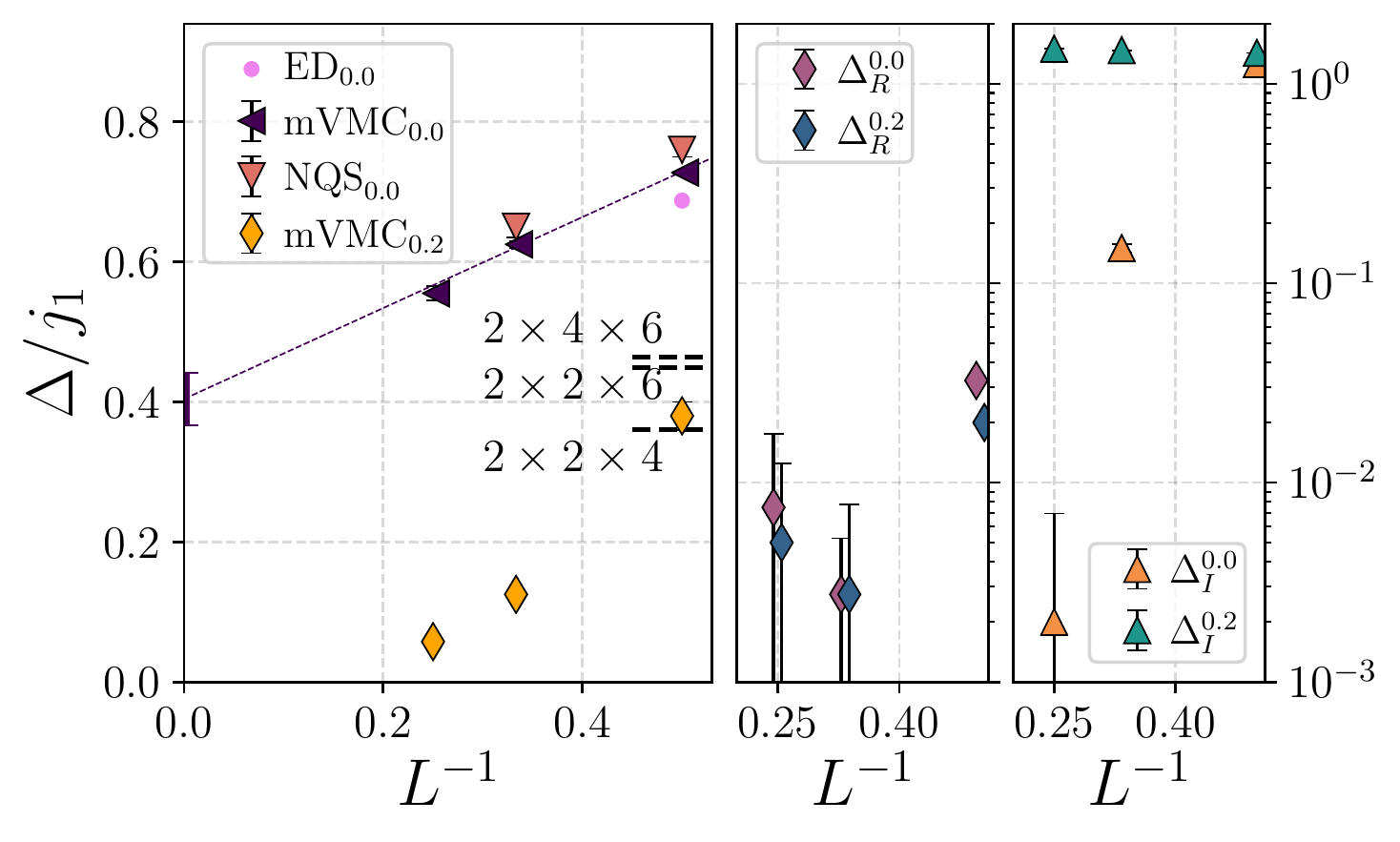}};
        \node at (-6.5, 2.75)  {(a) Spin gap};
        \node at (-2.2, 2.75)  {(b-c) Rotation/inversion};
    \end{tikzpicture}
    \caption{Spin/space-group spectroscopy at the $SU(2)$--symmetric point.\;(a) Triplet gap as a function of inverse linear system size $L$. Data points labelled $\mbox{ED}_{0.0}$, $\mbox{NQS}_{0.0}$ (difference between even and odd spin sectors projected using Eq.~\eqref{eq:parity_projection}), and $\mbox{mVMC}_{0.0}$ represent the spin gap at $j_2 / j_1 = 0.0$ obtained within ED, NQS, and mVMC, respectively, while $\mbox{mVMC}_{0.2}$ shows the vanishing spin gap at $j_2 / j_1 = 0.2$. Dashed lines on the right represent values obtained at non-equilateral clusters within mVMC at $j_2 / j_1 = 0.0$. The errorbar at $L^{-1} = 0$ represents $1/4$ of the difference between extrapolation and the value obtained on the largest possible cluster. The data can be found in Table\,\ref{tab:energies_spingaps} of Appendix\,\ref{sec:appendix_table}. (b-c)~Point-group symmetry gaps as a function of the inverse linear system size $L^{-1}$ obtained within mVMC. Data points labelled $\Delta_R^{0.0}$, $\Delta_I^{0.0}$ show the rotation and inversion gaps as $j_2 / j_1 = 0.0$, while $\Delta_R^{0.2}$, $\Delta_I^{0.2}$ stand for $j_2 / j_1 = 0.2$ point.}
    \label{fig:gaps}
\end{figure}

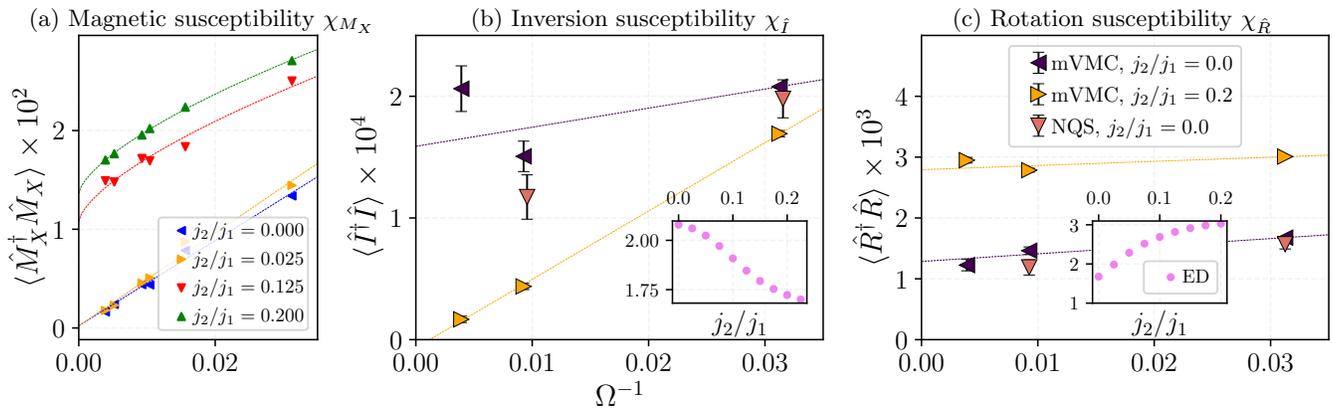
\begin{figure*}[t!]
    \centering
    \begin{tikzpicture}
        \node[inner sep=0pt] at (0.0, 0.0)    {\includegraphics[width=1.\textwidth]{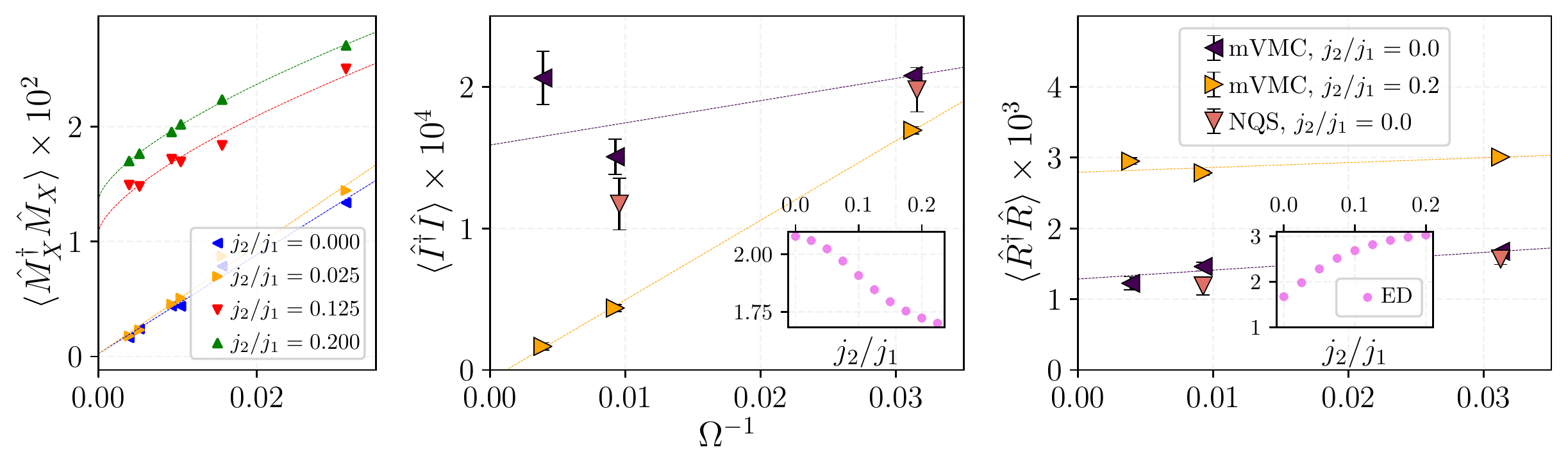}};
        \node at (-6.2, 2.7)  {(a) Magnetic susceptibility $\chi_{M_X}$};
        \node at (-0.5, 2.7)  {(b) Inversion susceptibility $\chi_{\hat I}$};
        \node at (5.9, 2.7)  {(c) Rotation susceptibility $\chi_{\hat R}$};
    \end{tikzpicture}
    \caption{Susceptibilities towards $SU(2)$, rotation and inversion symmetry breaking at $\lambda = 1$ point. (a)~Magnetic order susceptibility at $\mathbf k = X$ point extrapolated to infinite volume using $\Omega^{-1}$ scaling in the non-magnetic phase ($j_2 / j_1 = 0.0,\,0.025$) and $\Omega^{-2/3}$ in the ordered phase ($j_2 / j_1 = 0.125,\,0.2$). ({b})~The inversion-symmetry breaking susceptibility $\chi_{\hat I}$ as a function of inverse volume obtained within NQS and mVMC methods. The dashed lines show linear infinite volume extrapolation at $j_2 / j_1 = 0$ and $j_2 / j_1 = 0.2$. The inset shows the susceptibility dependence as a function of $j_2 / j_1$ obtained on a $4 \times 2^3$--cluster. ({c})~The rotation-symmetry breaking susceptibility $\chi_{\hat R}$ as a function of inverse volume obtained within NQS and mVMC methods. The dashed lines show linear infinite volume extrapolations at $j_2 / j_1 = 0$ and $0.2$. The inset shows the susceptibility dependence as a function of $j_2 / j_1$ obtained on a $4 \times 2^3$--cluster.}
    \label{fig:susceptibilities}
\end{figure*}

Taken together, our results strongly suggest that the phase boundary in the thermodynamic limit 
is located at substantially smaller $j_2 / j_1$
than the one extracted from $L=2$ ED and FRG, but remains at finite $j_2 / j_1$ in the region $\lambda \lesssim 1$. Since the initial ED phase boundary was well described by the parabolic fit, in Fig.~\ref{fig:intro}~(a) we also include the phase boundary uncertainty region as bounded by the two parabolic curves in the thermodynamic limit as a guide to the eye\footnote{Note that our data does not allow us to reliably draw conclusion about the shape of the phase boundary in the region $\lambda \ll 1$. The yellow region, shown in Fig.~\ref{fig:intro}~(a) at $\lambda \lesssim 0.8$ is a mere assumption.}. {We also point out that due to the shrinking width of the non-magnetic phase and its decreasing energy gain (as compared to the ordered $\mathbf k = \mathbf 0$ phase)\footnote{At $\lambda = 0$, already infinitesimal $j_2 / j_1 > 0$ orders the system to the $\mathbf k = \mathbf 0$ phase.}, variational approaches employed in this study converge to the ordered regime at $\lambda \leqslant 0.6$ upon performing the hysteresis technique on the $j_2 / j_1 = 0$ axis. Thus, we infer the non-magnetic region shape at $\lambda < 0.6$ only from the ED data presented in Appendix~\ref{sec:appendix_ED}.}

\subsection{Symmetry breaking in the non-magnetic phase}
\label{subsec:results_SSB}
We characterize the non-magnetic phase through its symmetry-breaking tendencies and contrast it with the behavior in the magnetic phase. We begin with the study of magnetic correlations and the spin gap, i.e., the energy difference between the lowest states in the $S = 0$ and $S = 1$ sectors. A vanishing spin gap allows for spontaneous spin-rotation symmetry breaking, characteristic of spin-nematic or magnetic phases. Numerous studies have already assessed the magnitude of the spin gap at $\lambda=1$, $j_2 / j_1 = 0$~\cite{doi:10.1143/JPSJ.67.4022,PhysRevLett.80.2933,doi:10.1143/JPSJ.70.640,hagymasi2020possible,PhysRevB.97.144407}. However, as seen in Table.~\ref{tab:energies_spingaps} of Appendix\,\ref{sec:appendix_table}, no definite conclusion can be made, since the results are sensitive to the specific cluster geometry~\cite{PhysRevB.97.144407}. Here, we obtain the spin gap on all available clusters, but, most importantly, on three  equilateral clusters $4 \times L^3$, which retain the pyrochlore point-group symmetries under consideration. We thus believe that our data provide the most reliable spin gap extrapolation to date. As is seen from Fig.~\ref{fig:gaps}~(a), spin-gap extrapolations dramatically differ in the magnetic and the non-magnetic phase, represented by parameter choices  $j_2 / j_1 = 0.2$ and $j_2 / j_1 = 0$, respectively. The gap extrapolates to zero and to a finite value $\Delta_S / j_1 = 0.40(4)$ in the magnetic and non-magnetic phases, respectively. The latter agrees with the recent DMRG result $0.36(3)$~\cite{hagymasi2020possible}. 

A vanishing spin gap allows to establish an $SU(2)$--breaking order parameter. We substantiate  this via an extrapolation of the magnetic susceptibility $\chi_{\hat M_{X}} = \langle \hat M^{\dagger}_X \hat M^{\phantom{\dag}}_X \rangle$ shown in Fig.~\ref{fig:susceptibilities}~(a). Since the phase-transition point $j_2 / j_1$ is size-dependent, at intermediate $j_2 / j_1$ no such extrapolation is possible for the system sizes available to us and we show only $j_2 / j_1$, which fall into one phase for all lattice volumes. For $j_2 / j_1 \leqslant 0.025$, $\chi_{\hat M_{X}}$ extrapolates to zero showing the absence of long-range magnetic correlations with this wave vector. {Here we employ the $\Omega^{-1}$ extrapolation as we expect spin-spin correlations to decay exponentially with distance in the non-magnetic phase~\cite{gong2014plaquette}. Note that scaling $\Omega^{-r}$ with $r < 1$ in the  non-magnetic phase (e.\,g. due to a large correlation length) could only strengthen the conclusion of vanishing magnetic order.} On the other hand, correlations at $j_2 / j_1 \geqslant 0.12$ extrapolate to finite values. {To show that, we use the $\Omega^{-2/3}$ scaling arising from gapless Goldstone mode contributions to long-range spin-spin correlation in 3D~\cite{PhysRevB.50.3877}.} The finite extrapolation values imply that the magnetic operator $\hat{M}_X$ will acquire a finite expectation value, $\langle \hat M_X \rangle \neq 0$, in the thermodynamic limit in the ordered $\mathbf k = \mathbf 0$ phase, while the expectation value vanishes in the non-magnetic phase.

Similarly to magnetic order and the corresponding $SU(2)$--symmetry breaking, we study the point-group symmetry breaking, which may be associated with dimer-type order, for instance~\cite{cite_1}. We compute the  energy difference between the ground state and the lowest state in sectors with a different inversion and rotation eigenvalue. Only the equilateral clusters with $4 \times L^3$ obey rotation symmetry. In Fig.~\ref{fig:gaps}~(b) we show these gaps in the two phases at $j_2 / j_1 = 0.0,\,0.2$ plotted against the inverse linear system size $L^{-1}$~\cite{nomura2020dirac,PhysRevLett.121.107202,PhysRevB.102.014417}.

The inversion-symmetry gap extrapolates to a finite value in the $\mathbf k = \mathbf 0$ phase, which is in agreement with the fixed-point wave function for this phase, since the inversion operator does not mix the pyrochlore sublattices (see Appendix~\ref{sec:appendix_breaking} for detailed description of magnetic phase and frustrated phase fixed-point wave functions).
In contrast, the inversion-symmetry gap vanishes in the non-magnetic phase, opening the prospect of spontaneous inversion symmetry breaking.

The rotation-symmetry gap between the ground state (eigenvalue 1) and the lowest rotational doublet (eigenvalues $e^{\pm 2\pi i/ 3}$) is found to be very small on the $4 \times 2^3$ cluster in both phases and falls below the error bar of our calculations in larger volumes, allowing for a spontaneous breaking of the rotation symmetry. In the $\mathbf k = \mathbf 0$ phase, this is in accordance with the sublattice order, which spontaneously breaks rotational symmetry. However, the rotation gap closing in the non-magnetic phase is a novel result. 

To further support these results, we compute the susceptibilities $\chi_{\hat I}$ and $\chi_{\hat R}$ to condensation of the operators
\begin{equation}
\label{eq:fixed_orders}
\begin{split}
    \hat I &= \hat{\mathcal{O}}( -1,  1), \\
    \hat R &= \hat{\mathcal{O}}\left( +1,  e^{2 \pi i / 3 }\right),
    \end{split}
\end{equation}
corresponding to the inversion pseudoscalar and rotation $E$ irreducible representation operators, where $\hat{\mathcal{O}}(\xi, \omega)$ is defined in Eq.~\eqref{eq:inversion_dimer}. In Fig.~\ref{fig:susceptibilities}~(b-c) we show the corresponding susceptibilities for both phases, corroborating the gap analysis. In the non-magnetic phase the infinite-volume extrapolations of $\chi_{\hat I}$ and $\chi_{\hat R}$ are non-vanishing and, notably, $\chi_{\hat R}$ is an order of magnitude larger than $\chi_{\hat I}$, which agrees with the relative gap magnitudes (see Appendix\,\ref{sec:appendix_ED}). Likewise, in the $\mathbf k = \mathbf 0$ phase, only $\chi_{\hat R}$ extrapolates to a nonzero value. {For extrapolation, we employ the $\Omega^{-1}$ scaling based on our expectation that, if the dimerized phase indeed stabilizes, it breaks no continuous symmetry and has no gapless modes that would lead to non-exponential long-range dimer-dimer correlation saturation (see Fig. 14 in Ref.~\onlinecite{PhysRevB.84.024406}), as it happens with spin-spin correlations. If, on the contrary, no dimer order is truly established, dimer-dimer correlations decay algebraically and may cause scaling other than $\Omega^{-1}$. Luckily, such change will lead to a scaling of the form $\Omega^{-r}$, with $r < 1$, which increases the slope against the one presented in the plots and thus further solidifies the conclusion. Real-space-resolved dimer-dimer correlations supporting this scenario are shown in Appendix\,\ref{sec:appendix_scaling}. We emphasize that if, despite this argument, the scaling dramatically differs from $\Omega^{-1}$, e.\,g., $L^{-1}$, this would not allow us to draw a conclusion about rotation susceptibility extrapolation in the non-magnetic phase shown in Fig.~\ref{fig:susceptibilities}~(c).} The insets of Fig.~\ref{fig:susceptibilities}~(b-c) show $\chi_{\hat I}$ and $\chi_{\hat R}$ obtained within ED. They show trends consistent with the larger-volume results.

\begin{figure*}[t!]
    \centering
    \begin{tikzpicture}
        \node[inner sep=0pt] at (0.0, 0.0)    {\includegraphics[width=0.9\textwidth]{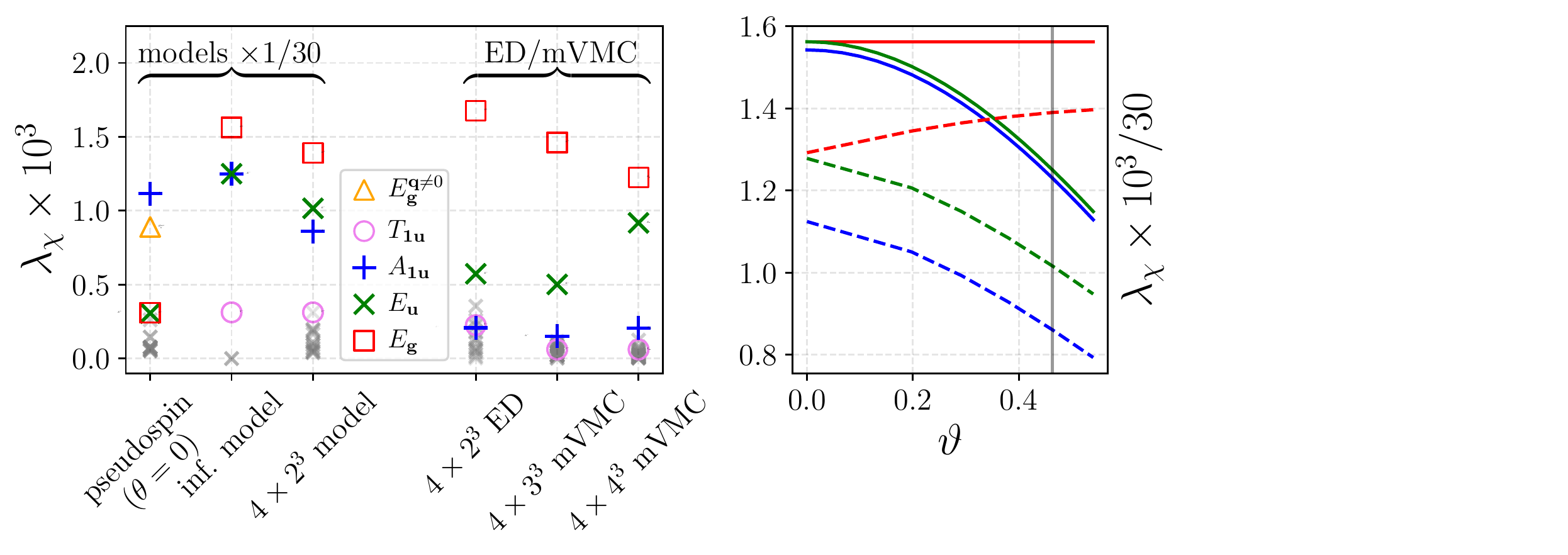}};
        \node[inner sep=0pt] at (6.3, 2.8)    {\includegraphics[width=0.12\textwidth]{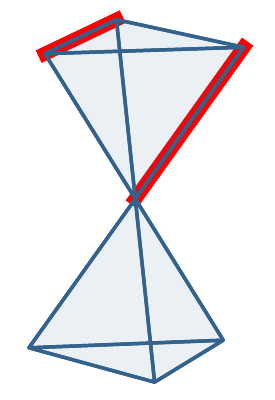}};
        \node[inner sep=0pt] at (8.3, 2.8)    {\includegraphics[width=0.12\textwidth]{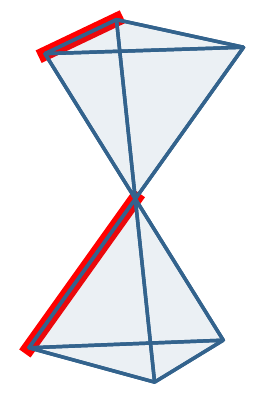}};
        \node[inner sep=0pt] at (7, -1.0)    {\includegraphics[width=0.27\textwidth]{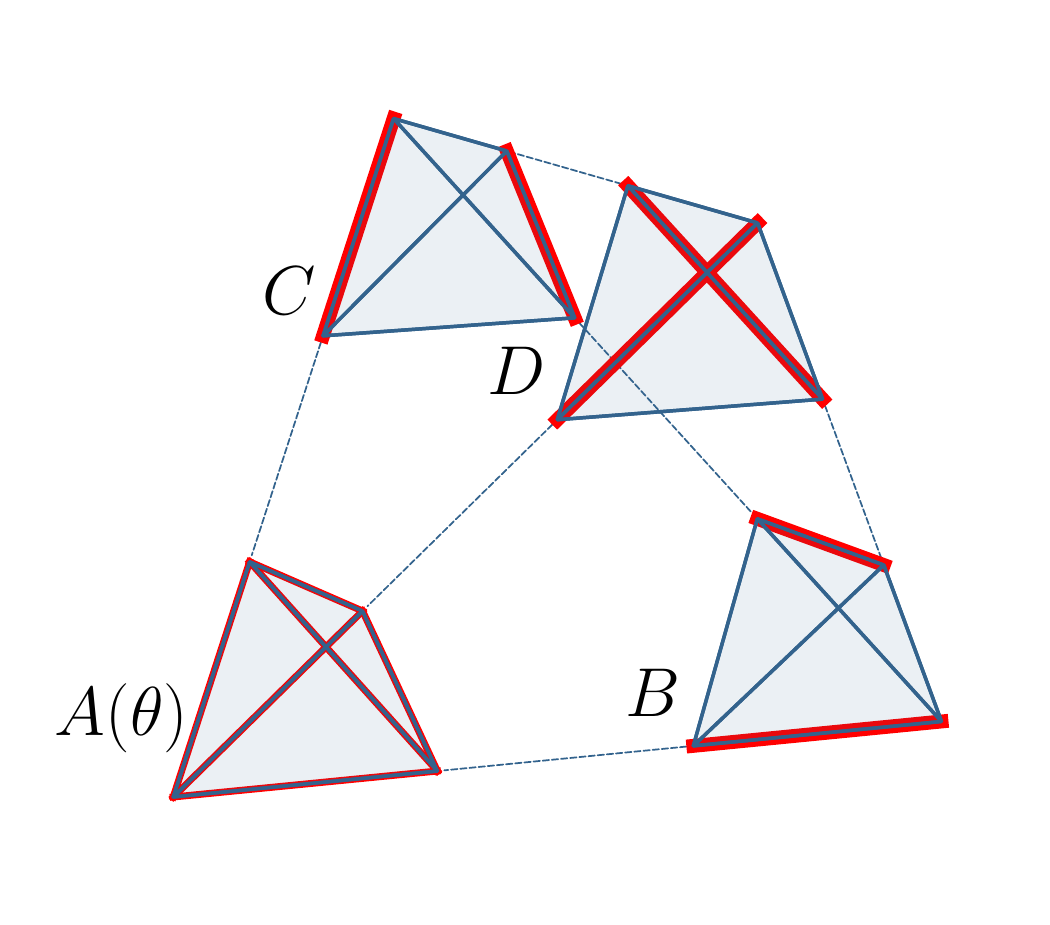}};
        \node at (-4.2, 2.8)  {(a) Spectroscopy: models and ED/mVMC};
        \node at (1.4, 2.8)  {(b) Eigenvalues flow};
        \node at (6.9, 4.6)  {(c) Dimerized wave function};
        \node at (6.9, 1.)  {(d) Pseudospin wave function};
    \end{tikzpicture}
    \caption{{(a)~Eigenvalues of the dimer-dimer correlation matrix and their irreducible representations of the point group computed at the $SU(2)$--symmetric point and $j_2 / j_1 = 0$. The grey crosses correspond to all eigenstates with small eigenvalue, irrespective of their irreducible representation, including all finite momentum sectors. The $E^{\mathbf q \neq \mathbf 0}_{\mathbf g}$ state is degenerate in the momentum sectors $(0,\,\pi,\,\pi),(\pi,\,0,\,\pi),(\pi,\,\pi,\,0)$. (b)~Evolution of eigenvalues ($ \times 1 / 30$) within the infinite-volume (continuous) and the $4 \times 2^3$-sites models (dashed) as the function of $\vartheta$. The colors coincide with the representation notations of (a). For the infinite-volume model, the blue line ($A_{1 \mathbf{u}}$) is slightly shifted for visibility as it always remains degenerate with $E_{\mathbf{u}}$. (c)~Fully dimerized $|\alpha\rangle$ (left) and $|\beta \rangle$ (right) wave functions. (d)~Pseudospin wave function introduced in Ref.~\cite{PhysRevB.65.024415}.}}
    \label{fig:eigenvalues_analysis}
\end{figure*}
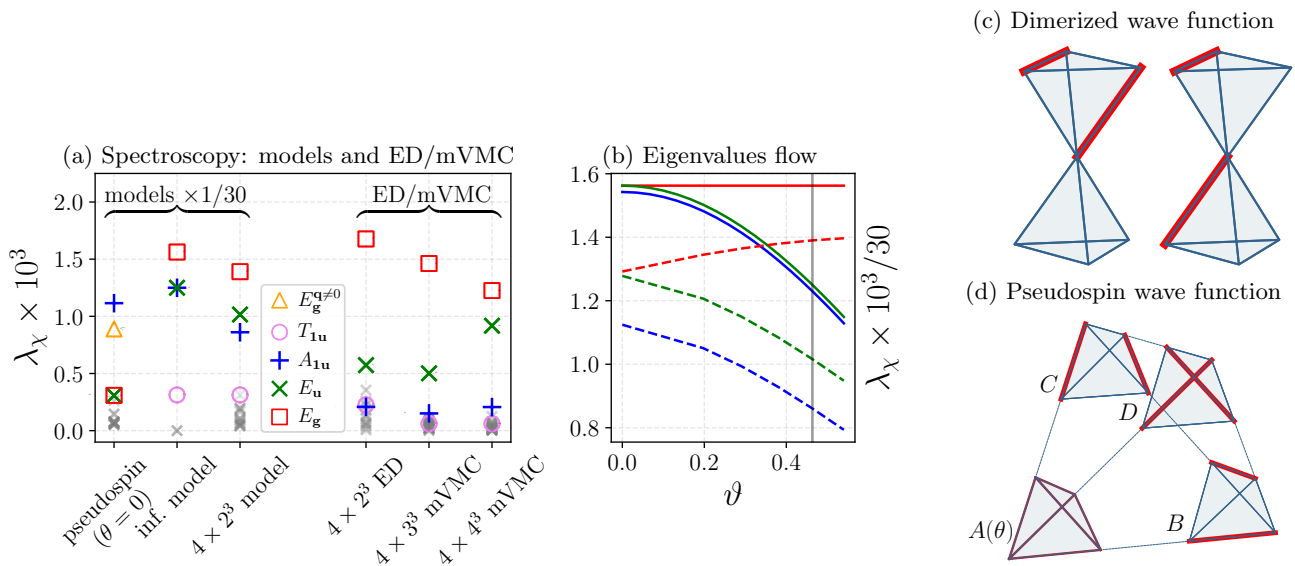

A summary of operator expectation values based on these results is shown in the frames in Fig.~\ref{fig:intro}~(a). Specifically, the rotation operator acquires a nonzero expectation value $\langle \hat R \rangle \neq 0$ in both phases, while the inversion operator is nonzero $\langle \hat I \rangle \neq 0$ only in the non-magnetic phase. 

{
\subsection{General dimerization pattern analysis}
\label{subsec:general_analysis}
Having observed non-vanishing susceptibilities towards establishment of rotation and inversion breaking dimerization patterns in the non-magnetic phase, we classify and measure (within ED/mVMC) all symmetry-allowed dimer observables, and compare them to those of analytically known model wave functions.

All the $12 \times L_x \times L_y \times L_z$ bonds of the pyrochlore lattice can be labeled with the unit cell index and an index within the unit cell. We will use the following convention: We choose each site of the 0--sublattice as the origin of a unit cell, where up- and down-tetrahedra meet, and assign the 12 bonds making up these two tetrahedra to the unit cell with the origin at site 0, as shown in Fig.~\ref{fig:intro}~(a). Given a quantum state, we then compute the dimerization tensor $\chi^D_{ij, \mu \nu} = \langle \hat D^{\mu}_i \hat D^{\nu}_j \rangle - \langle \hat D^{\mu}_i \rangle \langle \hat D^{\nu}_j \rangle$ with the indices $0 \leqslant i,\,j < L_x \times L_y \times L_z = \Omega / 4$ running over unit cells in the lattice and $0 \leqslant \mu,\,\nu < 12$ enumerating bonds within unit cells. The eigenvectors that correspond to the largest eigenvalues of $\chi^D$, represent the dominant correlation patterns developed within the state of interest.

The ground state obtained within the $4 \times 2^3$ ED study belongs to the $\mathbf k = \mathbf 0$ momentum sector and the trivial irreducible representation of the point-group symmetry, while on larger symmetric clusters $4 \times 3^3$ and $4 \times 4^3$ this property is enforced by the procedure described in Section~\ref{subsec:spwf}. Thus, $\chi^D_{ij, \mu \nu}$ shares the space group symmetries of the lattice (or finite cluster) and all its eigenstates transform according to an irreducible representation thereof. For all our numerically obtained variational wave functions, the Fourier transform of $\chi^D_{ij, \mu \nu}$ with respect to $i,\,j$ indices has its dominant eigenvalues in the $\mathbf q = \mathbf 0$ sector. Hence, the dominant dimer-dimer correlation patterns are translationally-invariant, unlike for instance for the $j_2/j_1$ Heisenberg model on the square lattice~\cite{nomura2020dirac}, where dimerization order breaks translational symmetry.

Restricting our consideration to translationally-invariant ($\mathbf q = \mathbf 0$) eigenvectors of $\chi^D_{\mu \nu}(\mathbf q = \mathbf 0)$, we classify them by the irreducible representations of the $O_h$ point group acting on the remaining 12-dimensional linear space, namely the $A_{1 \mathbf{g}},\,A_{1 \mathbf{u}},\,E_{\mathbf{g}},\,E_{\mathbf{u}},\,T_{\mathbf{1 g}}$ and $T_{\mathbf{1 u}}$ representations. The details about those representations and the action of symmetry operations can be found in Appendix\,\ref{sec:appendix_representations}. From the definition given in Eq.~\eqref{eq:fixed_orders}, we readily associate the $\hat I$ and $\hat R$ dimer operators, previously introduced in Eq.\,\eqref{eq:fixed_orders}, with the $A_{1 \mathbf{u}}$ and $E_{\mathbf{g}}$ irreducible representations, respectively.

In Fig.\,\ref{fig:eigenvalues_analysis}~(a) we show the dominant eigenvalues of $\chi^D_{\mu \nu}(\mathbf q = \mathbf 0)$ obtained from ED ($4 \times 2^3$ sites) and mVMC ($4 \times 3^3$ and $4 \times 4^3$ sites) states. Notably, the $E_{\mathbf{g}}$ irreducible representation is dominant. At the same time, we find as subleading eigenvector $E_{\mathbf{u}}$, which carries non-trivial quantum numbers of both rotation and inversion.
The appearance of $E_{\mathbf{g}}$ and $E_{\mathbf{u}}$ as the irreducible representations of the largest eigenvectors reaffirms our conclusions from Sec.\,\ref{subsec:results_SSB} that both rotation and inversion symmetry are likely broken in this phase.

Next, we want to connect the observed $\chi^D_{\mu \nu}(\mathbf q = \mathbf 0)$ eigenvalue hierarchy to a specific dimerized state. To this end, we consider the only two short-range translationally-invariant fully-dimerized patterns shown in Fig.\,\ref{fig:eigenvalues_analysis}~(c): two dimerized bonds in the same tetrahedron (denoted by $|\alpha \rangle$) or in opposite tetrahedra ($|\beta \rangle$). In general, the state may be a superposition of the two $|\psi(\vartheta)\rangle = \cos(\vartheta) \hat P |\alpha\rangle + \sin(\vartheta) \hat P |\beta \rangle$, which we parametrize by the angle $\vartheta$. To obtain a state that transforms trivially under all space group operations, as do our numerically obtained wave functions, we include the projector $\hat P$ discussed in Eq.\,\eqref{eq:q_projection}. 

In the infinite volume, the eigenvalues of $\chi^D_{\mu \nu}(\mathbf q = \mathbf 0)$ are computed exactly for this family of model wave functions. We repeat this analysis on the small $4 \times 2^3$ cluster where one can store the wave function directly. In Fig.\,\ref{fig:eigenvalues_analysis}~(b) we show eigenvalues of $\chi^D_{\mu \nu}(\mathbf q = \mathbf 0)$ as a function of $\vartheta$ for the finite and infinite volume. The eigenvalues (scaled to match the simulation data) are shown in Fig.\,\ref{fig:eigenvalues_analysis}~(a) for $\tan(\vartheta) = 1/2$ (this choice leads to a good qualitative agreement between models and numerical results and was chosen as intermediate value between $0$, where degeneracy between $E_{\mathbf{g}}$ and $E_{\mathbf{u}}$ is still present and $1$, where crossing with low-lying eigenstates happens). We point out good agreement in the dominant eigenvalues hierarchy and suggest $|\psi(\arctan(1/2))\rangle$ as a wave function qualitatively reproducing the numerical data. In this scenario, in the thermodynamic limit, when the point group symmetry is spontaneously broken, the wave function would be a superposition of the dimerization patterns $|\alpha\rangle$ and $|\beta\rangle$.

In addition, we discuss the pseudospin-ferromagnet wave function introduced, for instance, in Ref.~\cite{PhysRevB.65.024415}. Each isolated tetrahedron on the pyrochlore lattice has two orthogonal singlet states, which can be treated as states of the local pseudospin-$1/2$. We consider the pseudospin-ordered wave function introduced in Ref.~\cite{PhysRevB.65.024415}, where the tetrahedra are grouped into 4 meta-sublattices as shown in Fig.~\ref{fig:eigenvalues_analysis}~(d). We consider ferromagnetic pseudospin order in each of the sublattices. The pseudospin of the three sublattices (B, C, D) is obtained from the mean-field theory and results in a short-range dimerization pattern. The polarisation of A, expressed in terms of chiral superspin superpositions $|\pm \rangle$ as $(|+\rangle + e^{i \theta} |- \rangle) / \sqrt{2}$ is obtained by accounting for quantum corrections and depends on the phase $\theta$.  
In Fig.\,\ref{fig:eigenvalues_analysis}~(a) we show the eigenvalue hierarchy of the pseudospin wave function at $\theta = 0$ after projecting it with $\hat P$ to the trivial symmetry sector. Notably, the eigenstate structure is in sharp qualitative disagreement with our numerical data, for instance, due to appearance of $\mathbf q \neq \mathbf 0$ representation of $\chi^D_{ij, \mu \nu}$ among the leading contributions, which disfavours the pseudospin wave function scenario.}

\section{Discussion and Conclusion}
\label{sec:discussion}
The spin-1/2 Heisenberg model on the  pyrochlore lattice is an iconic candidate system for realizing a 3D QSL. This expectation is supported by the proximity of a $U(1)$ QSL in the vicinity of the Ising point and the absence of magnetic correlations at the Heisenberg point~\cite{iqbal2019quantum,PhysRevB.61.1149,PhysRevLett.80.2933,PhysRevB.79.144432}. 
To date, no consensus on the existence of the QSL phase has been reached, because 
the 3D many-body spin problem challenges all numerical techniques available. Confirming the QSL nature of a ground state is intrinsically harder than establishing an ordered phase:
besides proving the absence of long-range order, emergent fractionalized excitations should be found~\cite{Savary_2016}.

We have performed a comprehensive numerical search for spontaneous symmetry breaking in the putative QSL phase and quantified its extent in phase space. Combining two complementary variational approaches and exact diagonalization, we are able to gather strong evidence for a symmetry broken rather than a featureless QSL nature of the phase, based on extrapolations to the thermodynamic limit. 
We characterize the phase as not magnetically ordered, but spontaneously breaking rotation and inversion symmetry by establishing long-range dimer order. This contradicts the QSL realization at the Heisenberg spin-1/2 pyrochlore as discussed in Ref.~\cite{Savary_2016}. Rather, the symmetry breaking pattern for the nonmagnetic phase is consistent with the fixed-point dimerized wave function shown in Fig.\,\ref{fig:eigenvalues_analysis}~(c). Namely, simultaneous breaking of rotation and inversion symmetries by the dimer order would also be observed in the fully-dimerized state. In support of this, we found direct numerical evidence for dimer correlations in the ground state.

Its strong geometrical frustration  makes pyrochlore materials promising candidates for experimental observation of non-magnetic spin systems. For instance, spin-1 pyrochlore
$\mbox{NaCaNi}_2\mbox{F}_7$ with $j_2 / j_1 \sim -7 \times 10^{-3}$ was recently claimed to show spin-liquid-like behavior in neutron scattering experiments down to low temperatures~\cite{plumb2019continuum}. However, a spin-1/2 compound with nearly $SU(2)$--symmetric and dominantly nearest neighbor AFM interactions is still to be identified. 

We find that already a moderate admixture of  next-to-nearest-neighbor interactions establishes long-range magnetic order, emphasizing that only materials in a narrow parameter regime are expected to show non-magnetic ground states. Concretely, we showed that the width of this non-magnetic phase is $(j_2 / j_1)^* = 0.0295(30)$, which is substantially smaller than predicted in a recent FRG study~\cite{iqbal2019quantum}.

The seemingly small extend of parameter range is comparable to what is seen in other models with candidate QSL phases as well, in particular since it also extends to negative $j_2$. For instance, for the $j_2/j_1$ square Heisenberg model, the width of a putative QSL phase was found to be $\sim 0.05$~\cite{nomura2020dirac}.

Due to the shrinking of the non-magnetic phase as $\lambda$ is tuned from 1 to 0, the energy competition with the magnetic phase becomes even tighter and the local minima problem prevents us from obtaining a reliable non-magnetic variational wave function at small $\lambda$ and $j_2 / j_1 = 0$ for the $U(1)_{\pi}$ QSL. Severe finite size effects existing in this regime should also be noted: the $4 \times 2^3$ cluster hosts no $U(1)_{\pi}$ QSL state within perturbation theory on the spin-ice manifold as the $\lambda^3$-order hexagon flips are dominated by $\lambda^2$ terms specific to this cluster size~\cite{savary2012coulombic}. As a result, we were unable to locate the transition point between the observed symmetry-broken phase at $\lambda = 1$ and the well-known $U(1)_{\pi}$ QSL phase at $\lambda = 0$.
This transition, if it indeed exists, represents a crucial element of the pyrochlore puzzle that is yet to be resolved within future studies.

In summary, using many-variable Monte Carlo methods, we were able to refine the phase diagram of the spin-1/2 pyrochlore Heisenberg model both qualitatively and quantitatively, assembling strong evidence against a featureless QSL ground state. This still leaves open the possibility of a symmetry broken state with concomitant topological order.

\section{Acknowledgements}
We are sincerely grateful to A.\,G~Abanov, J.~Chang,  A.\,A.~Khudorozhkov, A.\,S.~Avdoshkin, A.\,A.~Bagrov, S.\,Sorella for useful discussions and to A.~Wietek for help on the early project stages. We thank F.~Becca and R.~Thomale for helpful comments on our manuscript. {Numerical simulations were performed using the open-source codes \texttt{mVMC}~\cite{misawa2019mvmc} for mVMC, \texttt{Netket}~\cite{netket:2019,doi:10.1143/JPSJ.77.114701} for NQS and \texttt{lattice{\_}symmetries}~\cite{westerhout2021latticesymmetries} for ED. } N.\,A is funded by the Swiss National Science Foundation, grant number: PP00P2{\_}176877 and the Simons Foundation. The work of T.\,W. was supported by European Research Council via Synergy Grant 854843 --- FASTCORR. A.\,T. is funded by the European Union’s
Horizon 2020 research and innovation program under the Marie Sklodowska Curie grant agreement No 701647. 
The authors acknowledge the use of the Flatiron computational facility, the Galileo supercomputer (CINECA) and computing resources of the federal collective usage center ``Complex for simulation and data processing for mega-science facilities'' at NRC ``Kurchatov Institute'', \href{http://ckp.nrcki.ru}{http://ckp.nrcki.ru}.

\appendix
\section{Neural Quantum States method and parameter training details}
\label{sec:appendix_NQS}
The NQS method casts each the spin configuration $\boldsymbol{\sigma}$ in $S^z$ basis to bit representation $\sigma_i \in \{-1, +1\}$ and uses it as an argument of a Neural Network $\Psi$. The result $\Psi(\boldsymbol{\sigma})$ is interpreted as the (not normalized) wave function component corresponding to the basis vector $\boldsymbol{\sigma}$. In case of a feed-forward neural network, which is most commonly applied in the NQS method, the input $\mathbf{v}^0 = \boldsymbol{\sigma}$ undergoes a sequence of transformations 
\begin{equation}
    \mathbf{v}^{i + 1} = f\left( \hat W_i \mathbf{v}^i + \mathbf{b}^i \right),
    \label{eq:dense}
\end{equation}
where $\hat W_i$ is the {\it weight matrix} and $\mathbf{b}^i$ is the {\it bias vector}. Parameters $(\hat W_i, \mathbf{b}^i)$ are the variational parameters of the NQS ansatz. The linear transformation $\hat W_i \mathbf{v}^i + b^i$ is followed by the application of nonlinearity $f$, which is an essential ingredient of the procedure, since otherwise only linear functions could be encoded. In this work, we apply the \texttt{ReLU} (rectified linear unit) nonlinearity~\cite{trabelsi2018deep} which is commonly used in artificial intelligence applications.

\begin{figure*}[t]
    \centering
    \includegraphics[scale=0.65]{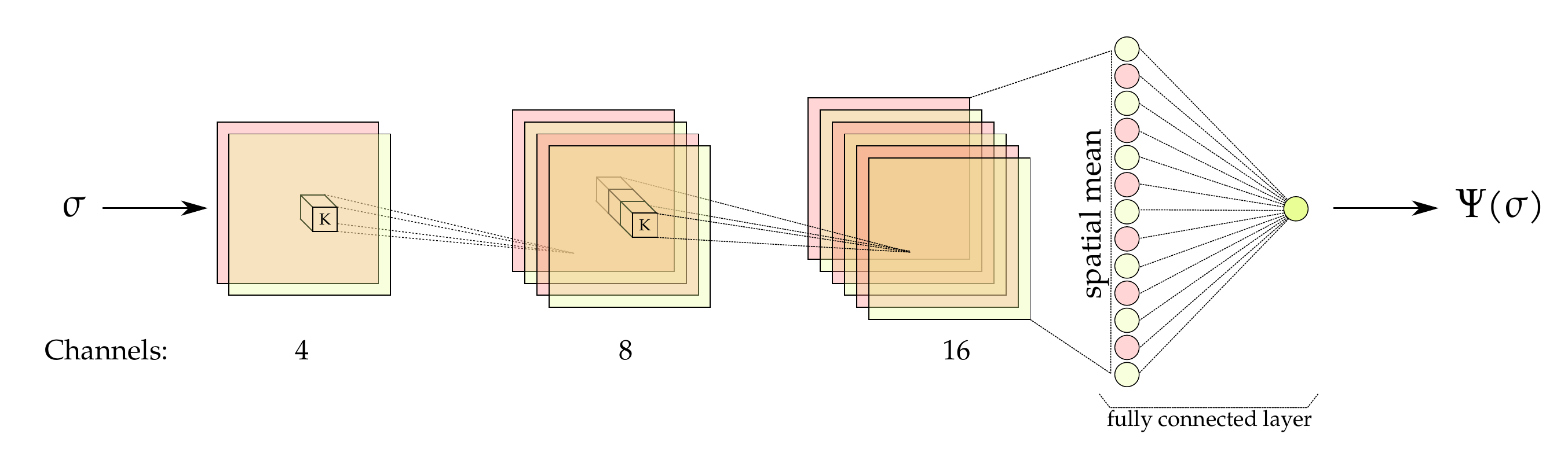}
    \caption{The standard translationally invariant neural network used in the paper. Spin configuration $\boldsymbol{\sigma} = \mathbf{v}^0_{l, \mathbf{r}}$ is first rearranged into the $(4, L_x, L_y, L_z)$ tensor, and then undergoes a sequence of convolutions in accordance with Eq.\,\eqref{eq:conv}. After the last convolution, the tensor $(C, L_x, L_y, L_z)$ is spatially averaged (in this example $C = 16$) and is fed to the fully connected layer, given by Eq.\,\eqref{eq:dense}. The resulting number is then interpreted either as $\log |\Psi(\boldsymbol{\sigma})|$ or $\mbox{arg} \Psi(\boldsymbol{\sigma})$.}
    \label{fig:my_label}
\end{figure*}

In NQS, fixing quantum numbers of the point symmetry group is done similarly to mVMC, using Eq.\,\eqref{eq:q_projection}. Imposing translational symmetry on the level of variational parameters requires construction of manifestly translation-invariant neural network. This ``hard-coded'' translational invariance turns out to be crucial for obtaining significant overlap with the QSL phase ground state~\cite{westerhout2019neural}, as it effectively increases the number of Monte Carlo samples $L_x \times L_y \times L_z$ times.

Translational invariance is achieved by using {\it convolutional layers} with {\it periodic padding}. Consider a spin configuration $\mathbf{v}^0_{l, x, y, z} = \boldsymbol{\sigma}_{l, x, y, z} \equiv \boldsymbol{\sigma}_{l, \mathbf{r}}$ to be rearranged in a 4D--tensor of shape $(C^0 = 4, L_x, L_y, L_z)$. Then, application of one convolutional layer reads
\begin{gather}
    \mathbf{v}^{i + 1}_{l, \mathbf{r}} = f \left(\sum\limits_{\Delta \mathbf{r}, l'} \mathcal{K}_{l l'}(\Delta \mathbf{r}) \mathbf{v}^{i}_{l', \mathbf{r} + \Delta \mathbf{r}}  + b^i_{l} \right).
    \label{eq:conv}
\end{gather}
Here, the {\it kernel} matrix $\mathcal{K}_{l l'}(\Delta \mathbf{r})$ depends only on the relative distance $\Delta \mathbf{r}$ between elements $\mathbf{v}^{i + 1}_{l, \mathbf{r}}$ and $\mathbf{v}^{i}_{l', \mathbf{r} + \Delta \mathbf{r}}$, while bias $b^i_{l}$ depends only on the sublattice index. If $\mathbf{r} + \Delta \mathbf{r}$ extends beyond the boundary, periodic boundary conditions are applied, which makes this architecture manifestly translation invariant. Note also that the output vector $\mathbf{v}^{i + 1}_{l, \mathbf{r}}$ has shape $(C^{i + 1}, L_x, L_y, L_z)$, so the number of channels $C^{i + 1}$ can vary while the spatial dimensions remain untouched.

As output the network should produce wave function in the form $\Phi(\boldsymbol{\sigma}) = (\log A, e^{i\phi})$, where $\log A$ is the amplitude logarithm and $e^{i\phi}$ is the complex phase of the wave function element. In this work, we train two separate neural networks, one of them producing amplitude, and the other one producing phase, which was shown to increase precision in a number of cases~\cite{doi:10.1002/qute.201800077,szabo2020neural}. So, the network should output one number obeying translational invariance. To construct one number from the hidden tensor of shape $(C, L_x, L_y, L_z)$, we first take the mean value over all spatial dimensions $(C, L_x, L_y, L_z) \to C$, and then apply a standard {\it dense layer} that maps $C$--dimensional vector to phase or amplitude (Eq.\,\eqref{eq:dense}).

Recently it has been shown that on moderately large systems, absence of ``hard-coded'' translational invariance might lead to outstanding wave function approximations~\cite{nomura2020helping}. Instead of CNN, one uses a standard shallow fully connected network with just one layer inspired by Restricted Boltzmann Machine (RBM) architecture used in early NQS studies~\cite{carleo2017solving}. In this way, no symmetries are encoded in the NN parameters, but rather all symmetry projectors, including momentum, are applied at the end. In this work, we also employ this architecture to improve the CNN results on the $4 \times 2^3$ and $4 \times 3^3$ systems.

Obtaining significant overlap with the QSL phase ground state is known to be a nontrivial task within the NQS method. The problem can be effectively reformulated in terms of the wave function {\it sign structure} (note that since the Hamiltonian Eq.\,\eqref{eq:hamiltonian} is real, phase of any element can be chosen $\phi = 0,\,\pi$). In an ordered phase, wave function element's sign can be easily inferred from the spin configuration $\boldsymbol{\sigma}$, e.g. via the Marshall--Peierls sign rule~\cite{doi:10.1098/rspa.1955.0200}. In this case, the network usually shows extreme accuracy, even outperforming existing approaches~\cite{carleo2017solving, PhysRevB.100.125124}. However, as the system is moved away from the ordered phase towards the QSL phase (for instance, by increasing $j_2 / j_1$ in the spin-1/2 Heisenberg model on square lattice), the overlap may drop to almost zero~\cite{szabo2020neural,PhysRevB.100.125124,westerhout2019neural}. It was shown that in the QSL phase the wave function phase structure has a much larger complexity~\cite{szabo2020neural}, which might be difficult to catch if a wrong training method or network architecture is applied. In case of 2D frustrated magnets it was shown that the right choice of network architecture can increase overlap with the ground state from 0 to $0.9$ at the maximally frustrated point~\cite{westerhout2019neural}.

In case of 3D frustrated magnets, the architecture alone turns out to be not enough to grasp the correct QSL ground state properties. To deal with frustrations, we introduce a novel algorithm of {\it alternating learning}. It was initially shown in~\cite{szabo2020neural} that one can improve the final training result by performing training in two stages. During the first stage, one sets amplitudes of all spin configurations equal $\log |\Phi(\boldsymbol{\sigma})| = 0$ and trains only phases. During the second stage, one trains both phase and amplitude networks simultaneously. We extend this idea and add the intermediate {\it alternating} phase, consisting of many tiny alternating trainings of phases with amplitudes fixed and amplitudes with phases fixed. 

Empirically, this extension can be motivated as follows. Note that within the NQS method the phase lives on the complex circle, while the true ground state phases can be defined purely real $\pm 1$. As the result, a converged network phase distribution usually has two major clusters at $\phi$ and $\phi + \pi$, where $\phi$ is an arbitrarily found physically irrelevant wave function gauge phase. After full convergence, the two phases are separated by a ``potential barrier'' and phase of a single spin configuration can not ``tunnel'' between these minima. However, if the amplitude landscape is totally flat, $\log |\Phi(\boldsymbol{\sigma})| = 0$, then the tunneling can happen with a higher probability. During the first training stage, we train only phases to find the best energy available with the constraint $\log |\Phi(\boldsymbol{\sigma})| = 0$. After this constraint is removed, amplitudes quickly converge and the potential barrier between $\phi$ and $\phi + \pi$ sets in. If, however, the amplitudes are trained only for a small time after which the phases are trained to adjust for slightly changed amplitudes, we allow some signs to ''tunnel`` between the two minima since the barrier is not yet too high.

\section{The hysteresis optimization}
\label{sec:appendix_hyst}
Ground states of neighboring phases show a pronouncedly different spin-spin susceptibilities behavior. In the vicinity of the phase transition, however, the two phases have a tight energy competition. Variational methods, being usually prone to ordered solutions, might get trapped in false minima and require a large number of random initial approximations to resolve the correct phase near the critical point. It turns out, however, that the search for the best variational parameters can be efficiently replaced with so-called ``hysteresis optimization''.

The first step is to optimize the wave function at two values of $j_2/j_1$ lying deep within the adjacent phases. The two sets of optimized wave function variational parameters are then used as the initial approximation for the {\it hysteresis procedure}. Namely, starting with the wave function trained deep within the $\mathbf{k} = \mathbf 0$ phase, we gradually decrease $j_2/j_1$, optimize the wave function for the new value of $j_2/j_1$, and obtain the energy $E_{\mathbf{k} = \mathbf 0}(j_2/j_1)$. During the evolution, we observe the susceptibility pattern $\chi_{\hat{M}_{\mathbf k}}$ to verify that the wave function still has the initial phase features. Repeating the same procedure in the opposite direction by starting with the QSL phase wave function, we obtain $E_{\mbox{\footnotesize QSL}}(j_2/j_1)$ that intersects $E_{\mathbf{k} = \mathbf 0}(j_2/j_1)$ at some $j_2/j_1 = (j_2/j_1)^*$, which provides us with an accurate estimate for the phase transition position. An example of the hysteresis optimization with mVMC for the largest $4 \times 4^3$ lattice is shown in Fig.~\ref{fig:corr_extrapolations}~(b). As the right panel shows, adiabatically evolved energies cross at $j_2 / j_1 \sim 0.032$. At this point, wave function of the magnetic phase starts to outperform the one of non-magnetic phase in terms of energy.  Importantly, at any $j_2/j_1$ the best energy obtained with the hysteresis optimization was never greater than the ones obtained with numerous trials starting from random approximations. Thus, the procedure provides us with a better energy in addition to saving the computational time. At every point we select the best available wave function and plot the magnetic order, which abruptly establishes if $j_2 / j_1$ is tuned above the phase transition point. This is consistent with the notion of metastable wave functions defined within magnetic and non-magnetic phases.

\section{Properties of the magnetic $\mathbf k = \mathbf 0$ phase}
\label{sec:appendix_k0}
\begin{figure}[b!]
    \centering
    \begin{tikzpicture}
        \node[inner sep=0pt] at (-6.5, 0)     {\includegraphics[width=0.40\columnwidth]{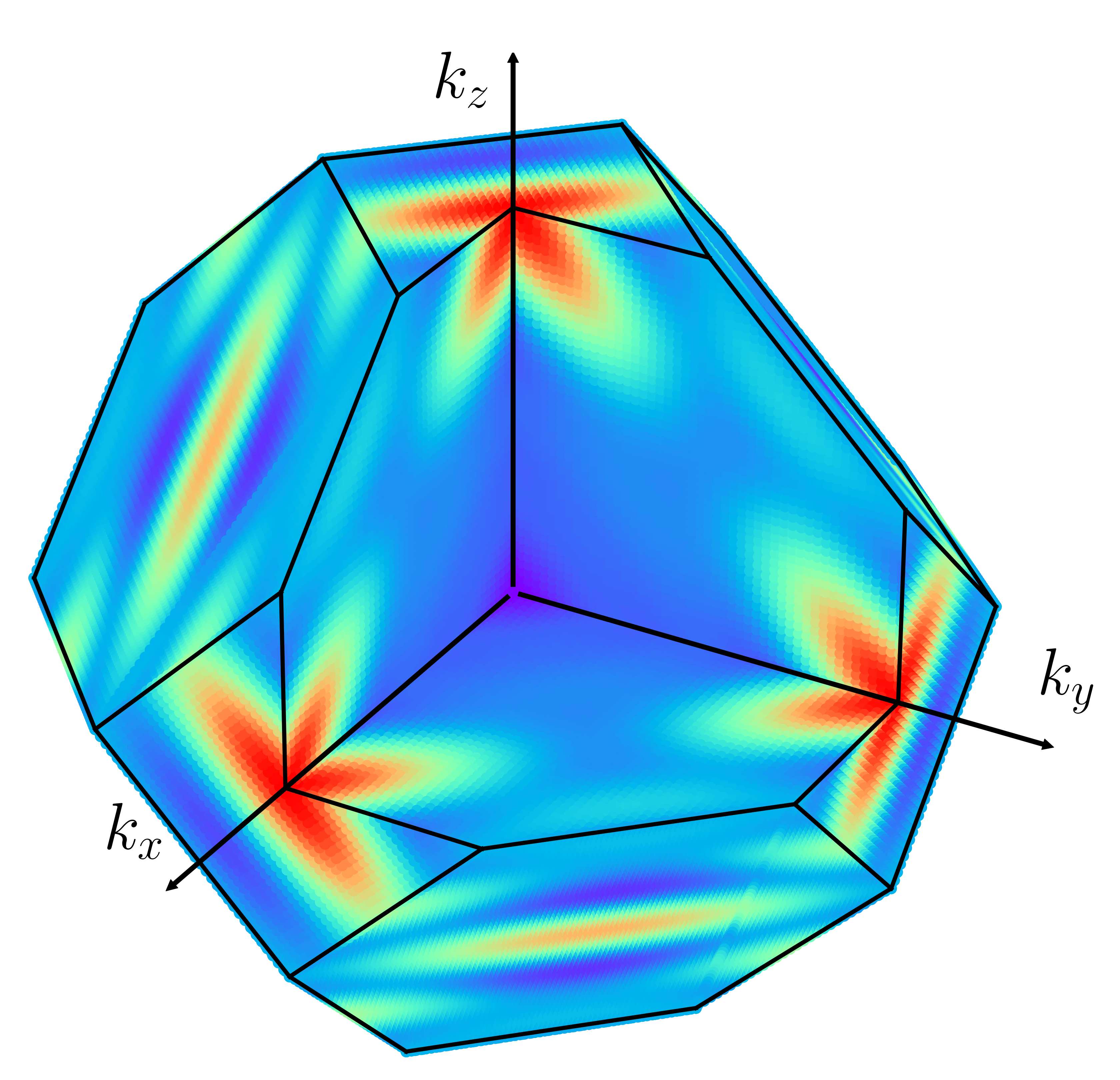}};
        \node[inner sep=0pt] at (-2.25, 0)     {\includegraphics[width=0.60\columnwidth]{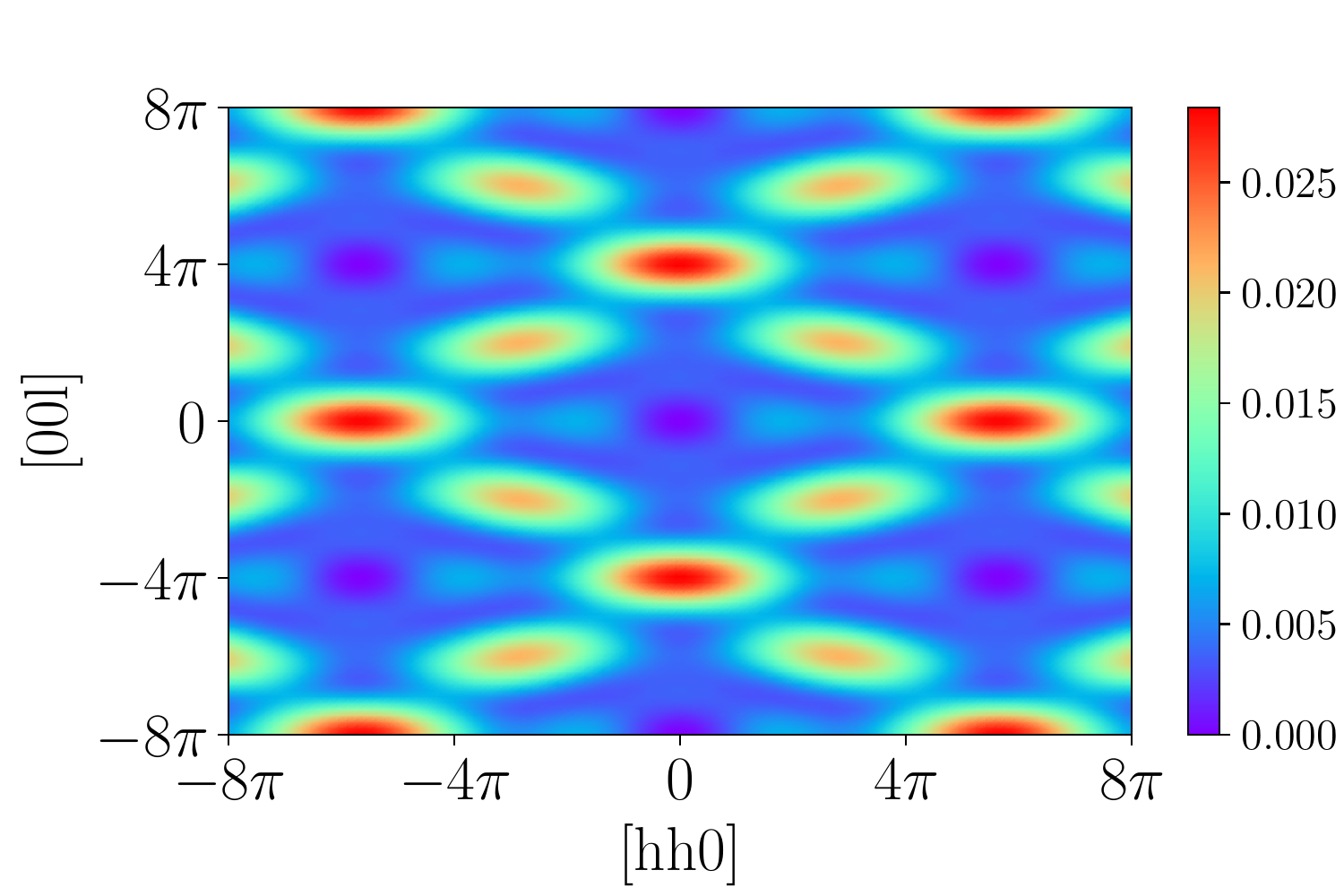}};
        \node at (-6.5, 1.75)  {\footnotesize (a) EBZ for $j_2 / j_1 = 0.2$};
        \node at (-2.25, 1.75)  {\footnotesize (b) $[hhl]$ cut for  $j_2 / j_1 = 0.2$};
        \node[inner sep=0pt] at (-6.5, -4.)     {\includegraphics[width=0.40\columnwidth]{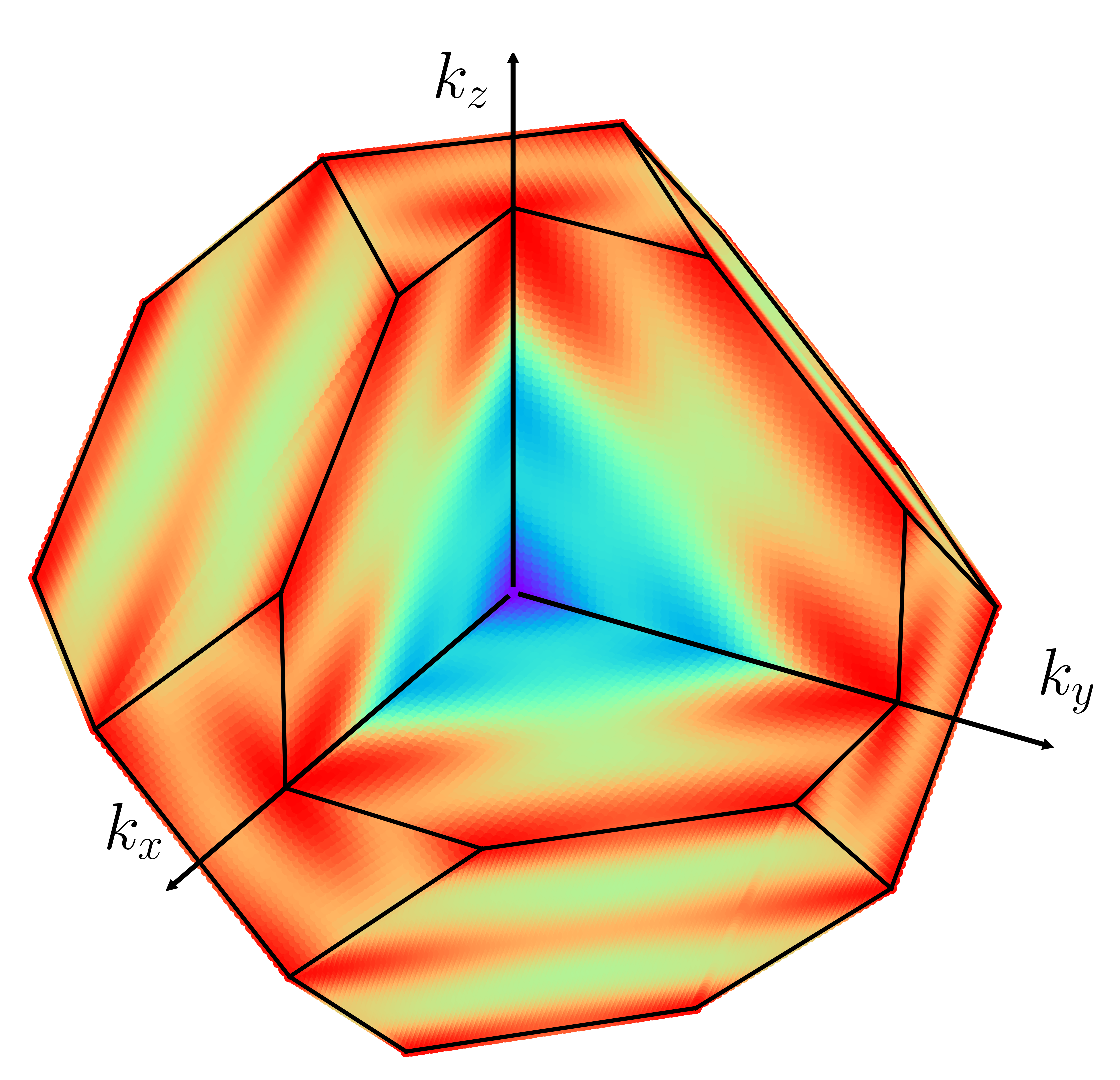}};
        \node[inner sep=0pt] at (-2.25, -4.)     {\includegraphics[width=0.60\columnwidth]{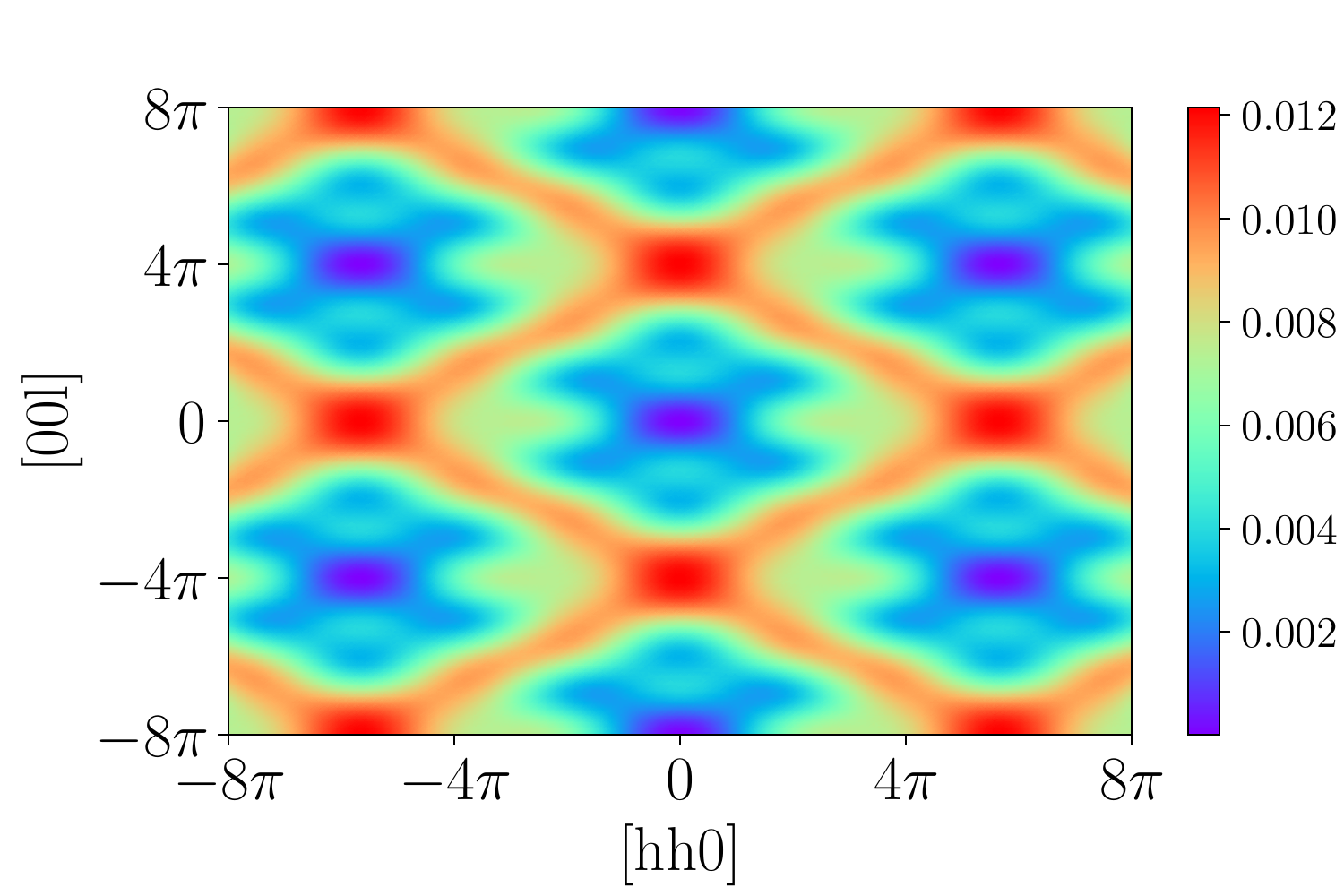}};
        \node at (-6.5, -2.3)  {\footnotesize (c) EBZ for $j_2 / j_1 = 0.0$};
        \node at (-2.25, -2.3)  {\footnotesize (d) $[hhl]$ cut for  $j_2 / j_1 = 0.0$};
    \end{tikzpicture}
    \caption{(a) Momentum-resolved correlations on the three-dimensional extended Brillouin zone obtained within ED on a $4 \times 2^3$ cluster with $\lambda = 1$, $j_2 / j_1 = 0.2$. (b) The $[hhl]$ plane slice of the extended Brillouin zone. (c-d) Same at $j_2 / j_1 = 0$.}
    \label{fig:corrs_hhl_qsl}
\end{figure}
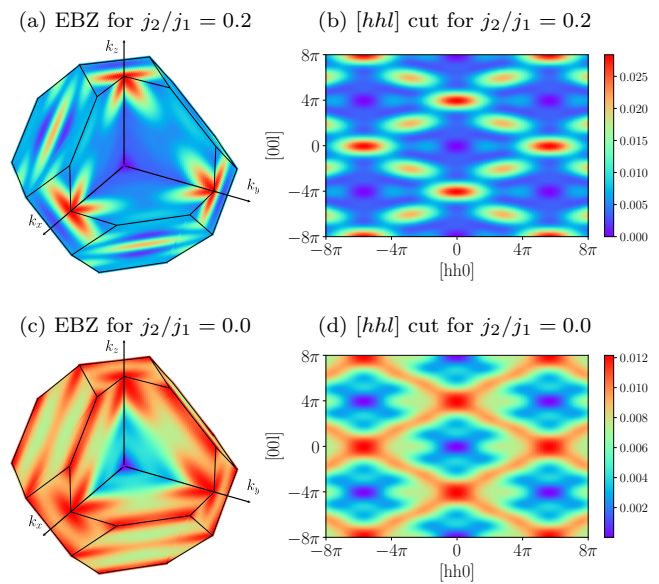

Momentum resolved spin-spin correlation functions are the most pronounced ordered phase fingerprints and thus are a powerful tool to track phase transition between ordered and disordered phases~\cite{iqbal2019quantum,gong2014plaquette,misawa2019mvmc,nomura2020dirac}. In terms of $\chi_{\hat{M}_{\mathbf k}} = \langle M^{\dagger}(\mathbf k) M(\mathbf k) \rangle$, the frustrated phase and the neighboring $\mathbf{k} = \mathbf 0$ ordered phase have several distinctive features. The phase, known also as the ``sublattice ferromagnet'', has spins within each sublattice ordered ferromagnetically, while still maintaining the zero-sum spin-ice rule within each tetrahedron. In case of perfect classical ordering, there are $C^2_4 = 6$ possible ground states satisfying this requirement. Each of these states can be tracked by two peaks of $\chi_{\hat{M}_{\mathbf k = X}}$ at a pair of inversion-related $X$--points (see Fig.~\ref{fig:corrs_hsl}). These states also have equally high peaks at all $L$--points (middle points of the large BZ faces) with the corresponding susceptibilities ratio $\chi_{\hat{M}_{\mathbf k = X}} / \chi_{\hat{M}_{\mathbf k = L}} = 4.$ In Fig.~\ref{fig:corrs_hhl_qsl}~(a-b) we show correlations magnitude over the 3D extended Brillouin zone and the $k_x = k_y$ cut called $[hhl]$.

\begin{figure*}[t!]
    \centering
    \begin{tikzpicture}
        \node[inner sep=0pt] at (0, 0.2)     {\includegraphics[width=0.98\textwidth]{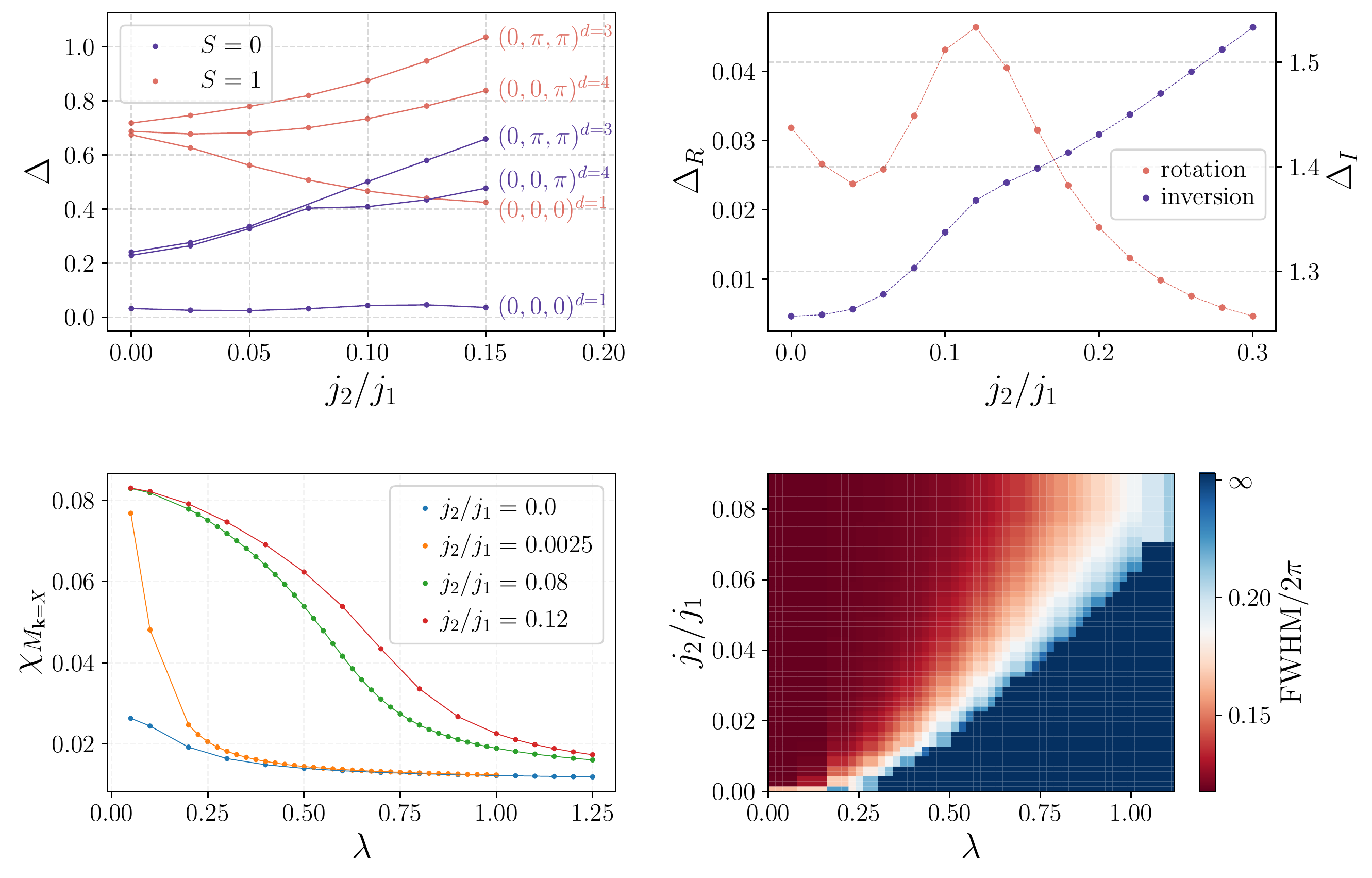}};
        \node at (-4, 5.9)  {\footnotesize (a) Spin/momentum spectroscopy within ED};
        \node at (-4,0.0)  {\footnotesize (c) Constant $j_2 / j_1$ slices};
        \node at (4, 5.9)  {\footnotesize (b) Gaps of $I / R$ on the $j_2/j_1$ axis within ED};
        \node at (4,0.0)  {\footnotesize (d) ED phase diagram};
    \end{tikzpicture}
    \caption{Results obtained within the ED study of the $4 \times 2^3$ cluster: level spectroscopy and phase diagram in terms of spin correlations.
        ({a}) The behavior of $E - E_{\mbox{\footnotesize GS}}$ spectrum for various quantum numbers as a function of $j_2 / j_1$ obtained within ED at $\lambda = 1$. The momentum $(0, 0, \pi)$ is always degenerate with $(\pi, \pi, \pi)$, thus we never show the latter. The upperscript $d$ denotes the total degeneracy of the energy level.
        ({b}) The gaps of rotation symmetry $\Delta_R$ and inversion symmetry $\Delta_{I}$ as functions of $j_2 / j_1$ obtained within the ED at $\lambda = 1$.
        ({c}) The peak height $\chi_{\hat{M}_{\mathbf k = X}}$ as a function of $\lambda$ for various values of $j_2 / j_1$. 
        ({d}) The peak $\chi_{\hat{M}_{\mathbf k = X}}$ half-width on the $(j_1 / j_1, \lambda)$ plane. If the peak width is undefined, then we put it equal infinity. 
    }
    \label{fig:ED_all}
\end{figure*}
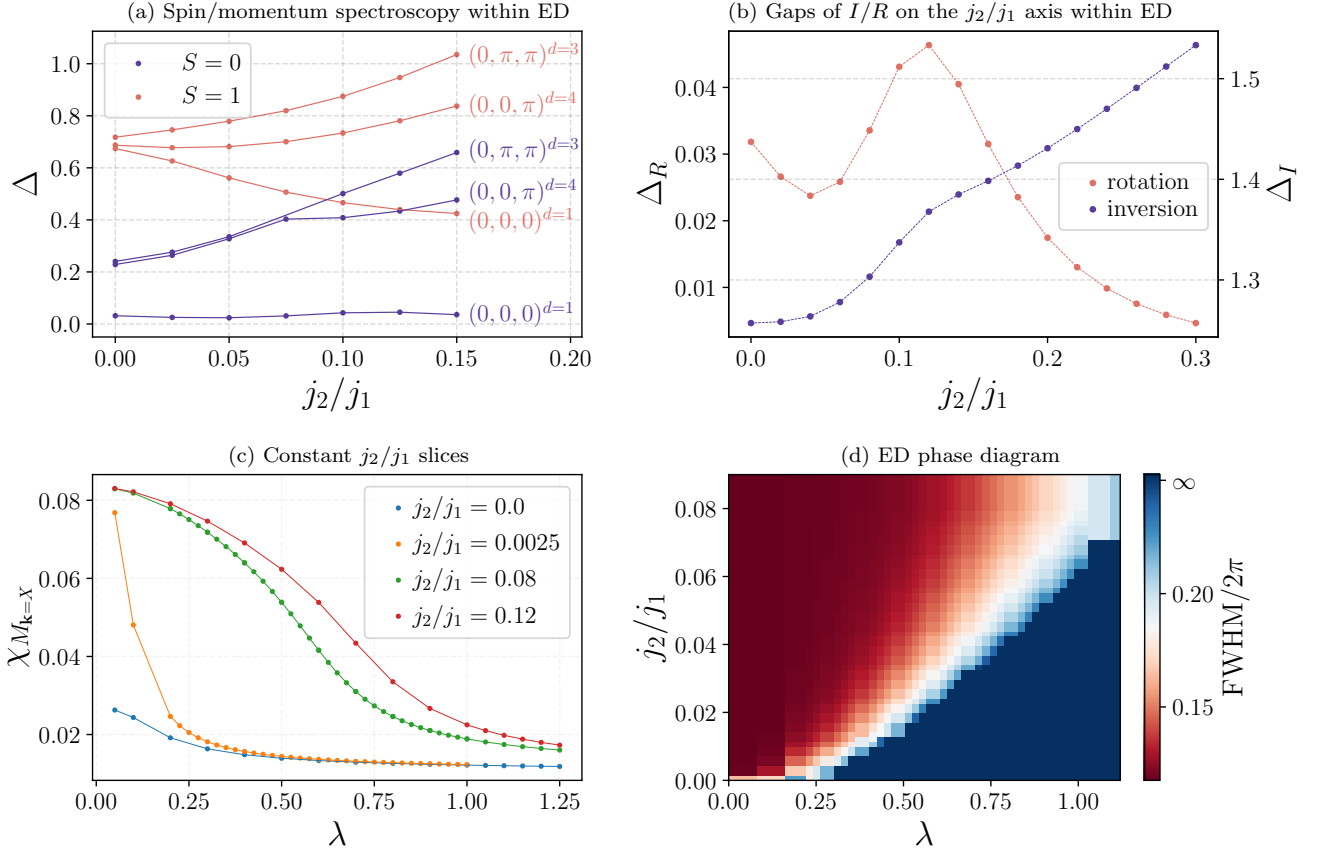

The frustrated phase, as compared to the $\mathbf k = \mathbf 0$ ordered phase, lacks any long-range magnetic order, $\chi_{\hat{M}_{\mathbf k}} = 0,\forall\,\mathbf{k}\,\in$\,BZ. The susceptibility multiplied by the system volume $\Omega \chi_{\hat{M}_{\mathbf k}}$, however, is finite and governed by the local spin-spin correlations. These correlations show the ``diffusive'' behavior, i.e. distributed over the BZ boundary (see Fig.~\ref{fig:corrs_hhl_qsl}~(c-d)). On the $[hhl]$ cut, the bow-tie pattern in the vicinity of $k_z = 4 \pi$ (vertical axis) is the attribute of the local spin-ice rule~\cite{iqbal2019quantum}.

\section{ED data: spectroscopy and phase diagram}
\label{sec:appendix_ED}
A qualitatively correct result for the shape of the phase diagram in terms of frustrated and magnetic phases can be obtained with ED on a $4 \times 2^3$ cluster. Beforehand, note that the $\mathbf k = \mathbf 0$ phase, though having qualitatively similar features on both $\lambda = 0$ and $\lambda = 1$ axes, such as magnetic susceptibility peaks at $X$ and $L$ points, has different peak heights. In the classical $\lambda = 0$ case, even a tiny positive $j_2 / j_1$ perfectly orders the system, giving rise to extreme $\chi_{\hat{M}_{\mathbf k = X}} = 1/12$. Contrary, in the quantum $\lambda = 1$ case, quantum corrections and $SU(2)$ symmetry do not allow perfect classical sublattice ferromagnetic ordering, which significantly reduces the peak. This is illustrated in Fig.~\ref{fig:ED_all}~(c), where we consider fixed--$j_2/j_1$ cuts of the phase diagram. Note that indeed if $j_2 / j_1 > 0$, the peak height $\chi_{\hat{M}_{\mathbf k = X}}$ would reach $1/12$, provided $\lambda$ was sufficiently small. Note also that as $\lambda > 1$, magnetic correlations are further reduced.

To unite these behaviors in the vicinity of $\lambda = 0$ and $\lambda = 1$ points, we consider the full width at half maximum (FWHM) characteristic defined as
\begin{equation}
    \chi_{\hat{M}_{\mathbf k = \mathbf X + \text{FWHM}}} = 1/2 \chi_{\hat{M}_{\mathbf k = X}},
\end{equation}
where $\mbox{FWHM}$ is a vector collinear to $\mathbf X - \mathbf W$. In Fig.~\ref{fig:ED_all}~(d) we show the peak width $|\mbox{FWHM}| / 2 \pi$ in the $(\lambda, j_2 / j_1)$ plane. If there is no such momentum $\mathbf k$ along the $XW$ line to satisfy the FWHM criterion, we put $|\mbox{FWHM}| / 2 \pi = \infty.$

Exact diagonalization provides another way to pinpoint phase transition between magnetically-ordered and frustrated phase by locating positions of level crossings between different total spin $S$ excitations~\cite{nomura2020dirac}. In Fig.~\ref{fig:ED_all}~(a) we show the energy gaps of excitations with distinct momenta and spin. 
Notably, the $\mathbf k = \mathbf 0$ excitation with $S = 1$ gradually softens and in the region $0.09 \leqslant j_2 / j_1 \leqslant 0.125$ it crosses non-zero momentum excitations of $S = 0$. This region may be taken for the spectroscopy estimation of the phase transition point, which is in rough agreement with the FWHM analysis. Note that softening of $S \neq 0$ mode is a signature of tendency to spontaneous $SU(2)$--symmetry breaking and stabilization of the magnetic phase.

Finally, in Fig.~\ref{fig:ED_all}~(b) we show inversion and rotation gaps as a function of $j_2 / j_1$ at $\lambda = 1$ within ED, that complements the susceptibility analysis in Section\,\ref{subsec:results_SSB}. Note that the inversion gap is significantly larger than the rotation gap, which explains why the susceptibility to rotation breaking was found to be of order of magnitude larger than the inversion breaking susceptibility. Also, the inversion gap decreases towards the non-magnetic phase, which hints developing tendency to inversion symmetry breaking, observed in Section\,\ref{subsec:results_SSB} within susceptibility analysis.

\section{Spontaneous symmetry breaking from ED data}
\label{sec:appendix_breaking}

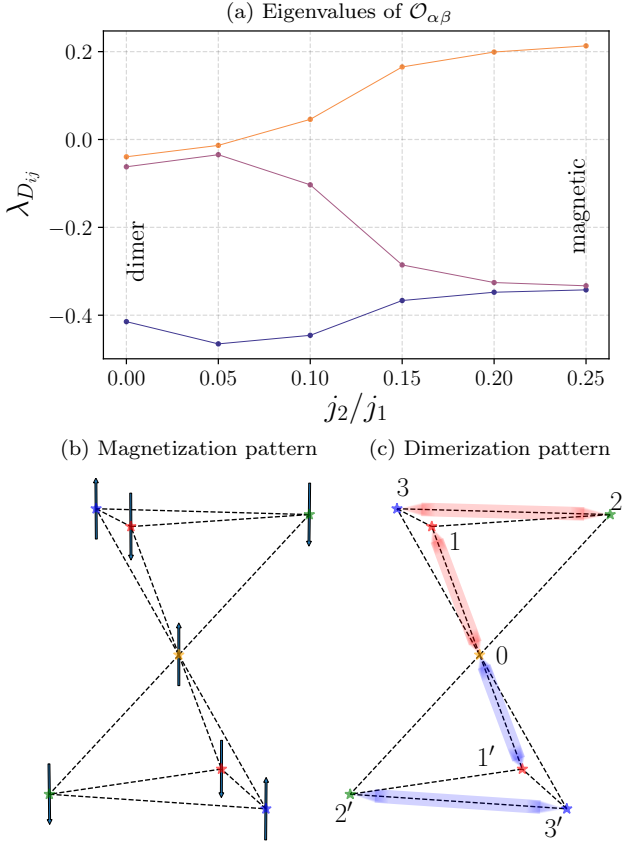
\begin{figure}[t!]
    \centering
    \begin{tikzpicture}
     \node[inner sep=0pt] at (0, 0)     {\includegraphics[width=0.5\textwidth]{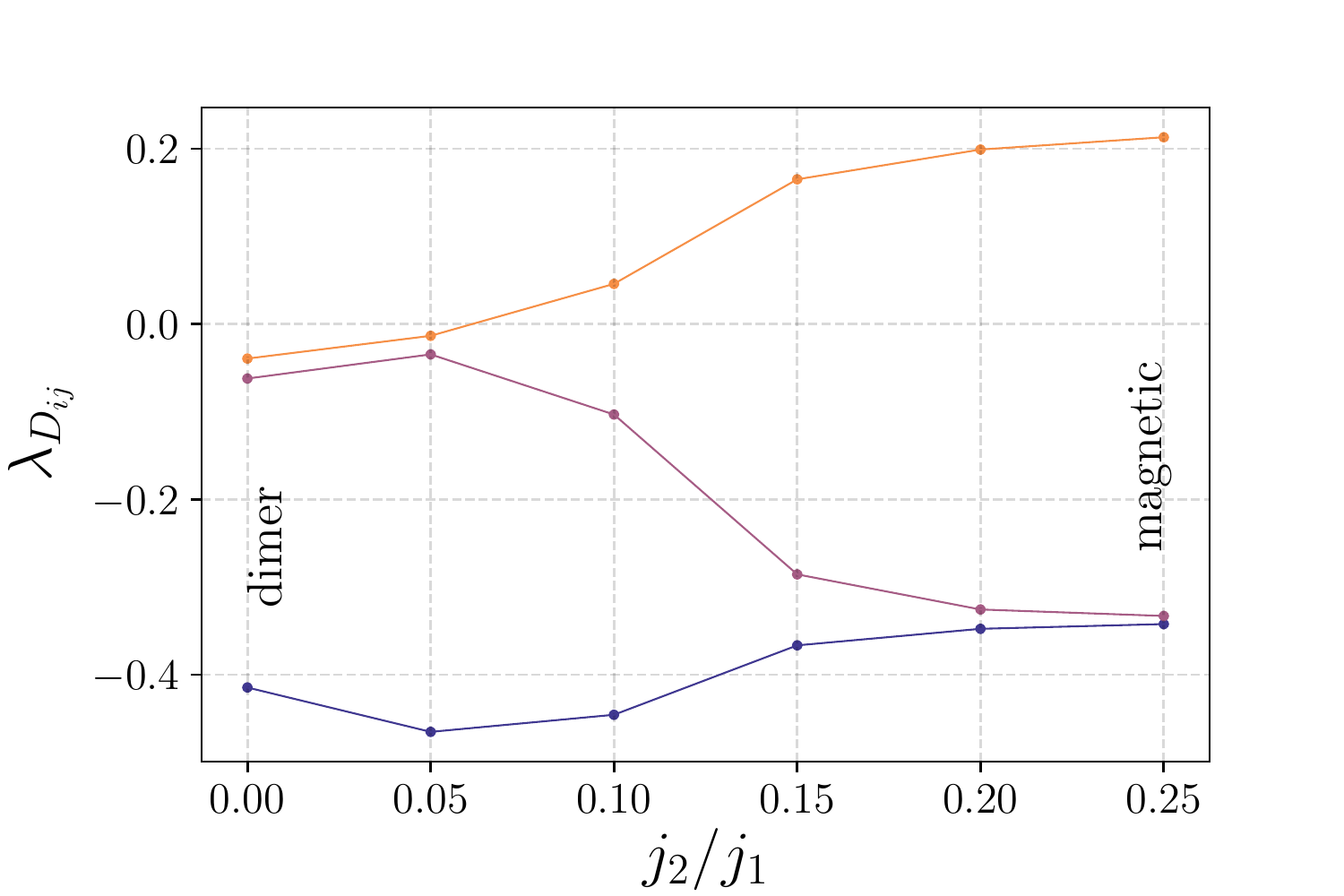}};
        \node[inner sep=0pt] at (-2, -6)     {\includegraphics[width=0.24\textwidth]{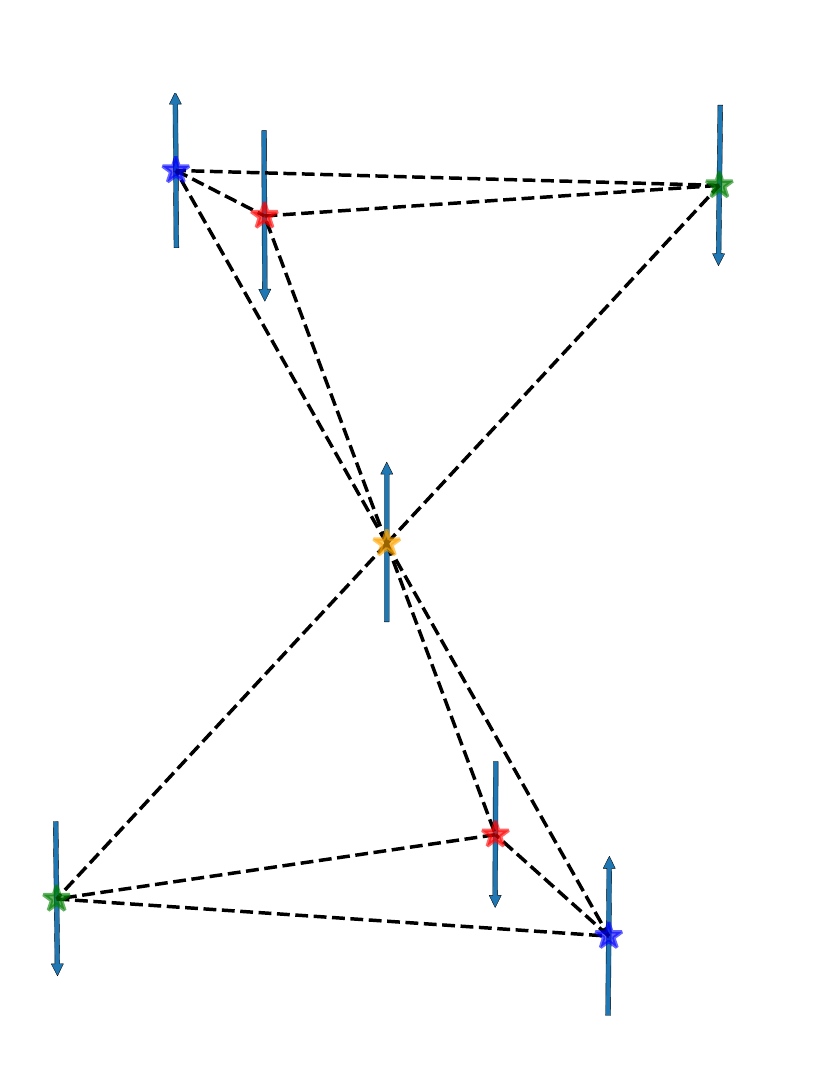}};
        \node[inner sep=0pt] at (2, -6)     {\includegraphics[width=0.24\textwidth]{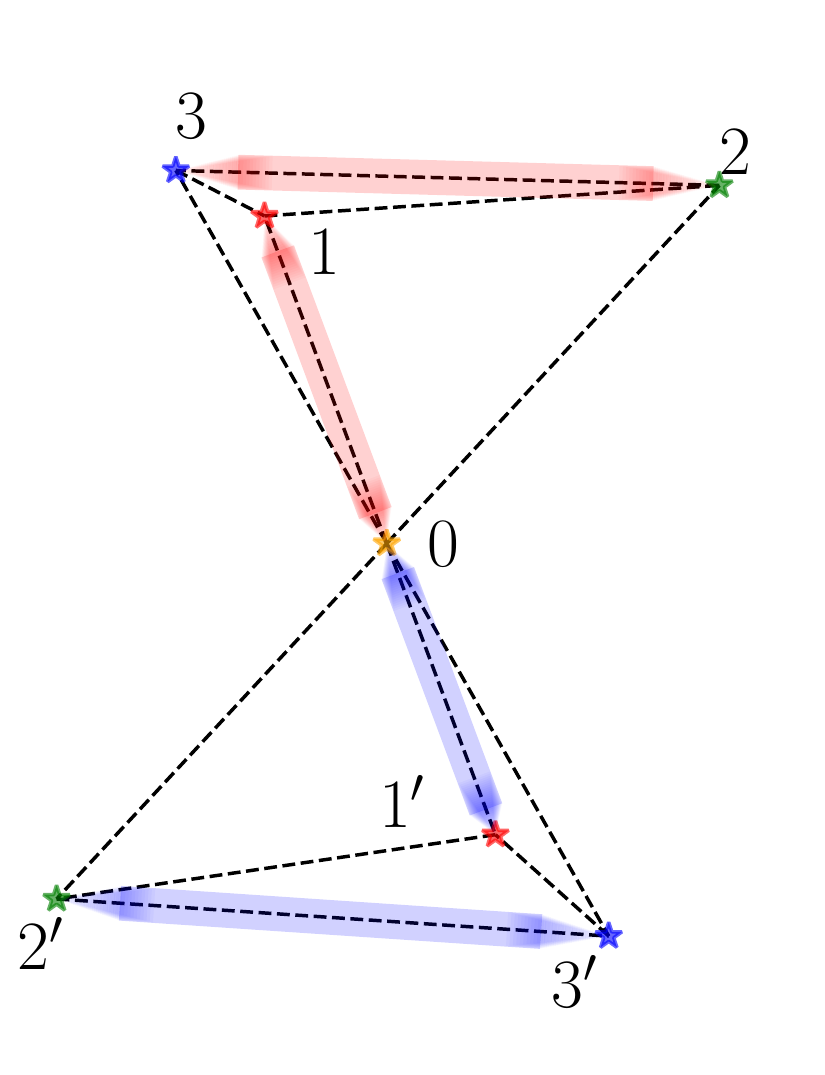}};
        \node at (0, 2.5)  {\footnotesize (a) Eigenvalues of $\mathcal{O}_{\alpha \beta}$};
        \node at (-2, -3.3)  {\footnotesize (b) Magnetization pattern};
        \node at (2, -3.3)  {\footnotesize (c) Dimerization pattern};
    \end{tikzpicture}
    \caption{({a}) Eigenvalues of $\mathcal{O}_{\alpha \beta}$ operator as a function of $j_2 / j_1$ { at $\lambda = 1$}. The ``dimer'' and ``magnetic'' labels denote the regions where the behavior is well-described with dimer and magnetic fixed point wave functions respectively.\;({b}) Tetrahedron unit cell showing one of the 6 classical magnetic $\mathbf k = \mathbf 0$ states ordered in each sublattice. Sublattices are shown in different colors. \;({c}) A tetrahedron unit cell showing one of the 3 dimerization patterns. Red and the blue colors represent two patterns related by inversion symmetry.}
    \label{fig:dimerisation_ed}
\end{figure}

In Fig.~\ref{fig:ED_all}~(b) we show that even on the ED--accessible $4 \times 2^3$ cluster the rotation gap between the ground state $s$--orbital and first excited $p_x / p_y$--orbitals is vanishingly small. As shown in Section\,\ref{subsec:results_SSB} this gap vanishes in the thermodynamic limit. 

To get more insight we apply the procedure suggested in~\cite{PhysRevB.88.245123}.
Assuming that these three low-energy states are ``already degenerate'', we can linearly superpose them at no energy cost. To probe the symmetry breaking, we select a bond dimer operator $\hat D_{ij} = \hat{\mathbf S}_i \cdot \hat{\mathbf S}_j$, residing on the fixed bond $A B'$ (see Fig.~\ref{fig:dimerisation_ed}~(c)), and study how its expectation may change as the ground states are superposed.

Within ED, we measure the $3 \times 3$ pair-wise expectation value matrix 
\begin{equation}
    \hat{\mathcal{O}}_{\alpha \beta} = \langle \psi_{\alpha} | \hat D_{ij} | \psi_{\beta} \rangle,
\end{equation}
where $0 \leqslant \alpha, \beta \leqslant 2$ index the approximately-degenerate ground states and the two excited states. Eigenvalues $\lambda_{D_{ij}}$ of $\mathcal{O}_{\alpha \beta}$ stand for the $\hat D_{ij}$ expectation values on superpositions where $\hat D_{ij}$ is diagonal. We remind here that the doublet gap only becomes smaller as $j_2 / j_1$ is increased to positive values, as seen in Fig.~\ref{fig:ED_all}~(b).

In Fig.~\ref{fig:dimerisation_ed}~(a) we show $\lambda_{D_{ij}}$ eigenvalues as a function of $j_2 / j_1$. Note that a nearly-degenerate pair of $\lambda_{D_{ij}}$ splits in the limit of large $j_2 / j_1$ and the intermediate eigenvalue switches to become near-degenerate with the upper eigenvalue. This dramatic rearrangement of $\lambda_{D_{ij}}$ is a yet another signature of a phase transition. To get more insight, we describe this behavior in terms of fixed-point classical wave functions. At large $j_2 / j_1$, when the system is in the $\mathbf k = \mathbf 0$ phase, the classical ground state has each sublattice ordered ferromagnetically, maintaining the zero spin per tetrahedron. In total, there are $4! / (2! 2!) = 6$ ways to construct such ground state. An example of such ground state is shown in Fig.~\ref{fig:dimerisation_ed}~(b). Since our case is not classical and the $SU(2)$--symmetry is present, we enforce the symmetry $s_z \to -s_z$ by turning $6$ such ground states into $3$ equal-weight superpositions symmetric under $s_z \to -s_z$ spin flip. Each state of this basis breaks rotation symmetry. The $\hat D_{ij}$ bond operator is diagonal in this basis and its expectation equals $\pm 1/4$ if the adjacent spins are (non-)collinear. Thus, the eigenvalues are $4 \lambda_{D_{ij}} = \{+1/4, -1/4, -1/4\}$, which is in a striking agreement with the right side (magnetic part) of Fig.~\ref{fig:dimerisation_ed}~(a).

In the other limit, $j_2 / j_1 \to 0$, where magnetic ordering is absent, the levels picture can be well described by the dimerized pattern breaking rotation symmetry. An example of such pattern is shown in Fig.~\ref{fig:dimerisation_ed}~(c). In a conventional dimerization pattern, each spin belongs only to one dimer. In our case, since the inversion symmetry is preserved (we only superpose inversion-symmetric states), a ground state must be taken as an equal-weight superposition of the red pattern and blue patterns, dimerizing up-tetrahedra or down-tetrahedra respectively and related by the inversion symmetry. This pattern can be rotated, producing three distinct ground states, each of them breaking rotational symmetry. The dimer operator is diagonal in this basis as well, and has the eigenvalues read $\lambda_{D_{ij}} = \{-3/8, 0, 0\}$, which is again in qualitative and even quantitative agreement with the left side (dimer part) of Fig.~\ref{fig:dimerisation_ed}~(a). 

Note, that we by no means claim that these tentative magnetized and dimer states are the ground states of the true system. For instance, the dimerized pattern has a much lower energy and only a 30\,\% overlap with the ground state. However, we note that the observed eigenvalues $\lambda_{D_{ij}}$ evolution picture is in acute agreement with this fixed point rotation-breaking wave functions treatment. We thus conclude that, since the rotation doublet gap closes in thermodynamic limit, the ground state may easily form a superposition with a strongly broken rotation symmetry, where the word ``strongly'' means that the magnitude of the symmetry-breaking operator $\hat D_{ij}$ is in striking agreement with the very typical magnetic and dimer symmetry-breaking states.

{
\section{Irreducible representations of dimer orders}
\label{sec:appendix_representations}
In this appendix, we illustrate the action of symmetry operations on $\mathbf q = \mathbf 0$ translationally-invariant dimer coverings and list irreducible representations in this 12-dimensional linear space. A vector in this space represents a complex amplitude choice on each of the $\{01,\,02,\,03,\,23,\,31,\,12,$ $01',\,02',\,03',\,2'3',\,3'1',\,1'2'\}$ bonds (see Fig.\,\ref{fig:dimerisation_ed}~(c) for definition of bond labeling). The action of symmetry generators on this space is given by 

\begin{align}
    C_3:\; &\{01 \to 02 \to 03 \to 04\},  \\
           &\{23 \to 31 \to 12 \to 23\}, \nonumber \\
           &\{01' \to 02' \to 03' \to 01'\}, \nonumber \\
           &\{2'3' \to 3'1' \to 1'2' \to 2'3'\}; \nonumber
\end{align}

\begin{align}
    I:\;   &\{01 \leftrightarrow 01'\}, \{02 \leftrightarrow 02'\}, \{03 \leftrightarrow 03'\}, \\
           &\{23 \leftrightarrow 2'3'\}, \{31 \leftrightarrow 3'1'\}, \{12 \leftrightarrow 1'2'\}; \nonumber
\end{align}

\begin{align}
    M:\;   &\{02 \leftrightarrow 12\}, \{03 \leftrightarrow 31\} \\
           &\{02' \leftrightarrow 1'2'\}, \{03' \leftrightarrow 3'1'\}, \nonumber
\end{align}
where $C_3$ is the three-fold rotation with respect to the ``easy axis'', $I$ is the inversion with respect to the $0$ point and $M$ is the mirror with respect to the plane passing through the bond 32 and the middle of 01. Note that those symmetry operations can map bonds from one unit cell to another. However, since we assume translational invariance of dimer pattern, which makes unit cells equivalent, we show only intra-unit-cell index of mapping and omit unit cell index. The resulting irreducible representations are listed in Table\,\ref{tab:dimer_irreps}.

\begin{table}[t!]
    \centering
    \scalebox{0.88}{
    \begin{tabular}{|c|c|c|c|c|c|c|c|c|c|c|c|c|}
    \hline
     repr. & 01 & 02 & 03 & 23 & 31 & 12 & 01' & 02' & 03' & 2'3' & 3'1' & 1'2' \\ \hline
     $A^{\mathbf{g/u}}$  & 1 & 1 & 1 & 1 & 1 & 1 & $\lambda$ & $\lambda$ & $\lambda$ & $\lambda$ & $\lambda$ & $\lambda$ \\
     \hline
     $E_+^{\mathbf{g/u}}$ & 1 & $\omega$ & $\omega^*$ & 1 & $\omega$ & $\omega^*$ & $\lambda$ & $\omega \lambda$ & $\omega^* \lambda$ & $\lambda$ & $\omega \lambda$ & $\omega^* \lambda$ \\ \hline
     $E_-^{\mathbf{g/u}}$ & 1 & $\omega^*$ & $\omega$ & 1 & $\omega^*$ & $\omega$ & $\lambda$ & $\omega^* \lambda$ & $\omega \lambda$ & $\lambda$ & $\omega^* \lambda$ & $\omega \lambda$ \\ \hline
     $T_a^{\mathbf{g/u}}$ & 1 & 1 & 1 & -1 & -1 & -1 & $\lambda$ & $\lambda$ & $\lambda$ & $-\lambda$ & $-\lambda$ & $-\lambda$ \\
     \hline
     $T_b^{\mathbf{g/u}}$ & 1 & $\omega$ & $\omega^*$ & -1 & $-\omega$ & $-\omega^*$ & $\lambda$ & $\omega \lambda$ & $\omega^* \lambda$ & $-\lambda$ & $-\omega \lambda$ & $-\omega^* \lambda$ \\ \hline
     $T_c^{\mathbf{g/u}}$ & 1 & $\omega^*$ & $\omega$ & -1 & $-\omega^*$ & $-\omega$ & $\lambda$ & $\omega^* \lambda$ & $\omega \lambda$ & $-\lambda$ & $-\omega^* \lambda$ & $-\omega \lambda$ \\ \hline
    \end{tabular}
    }
    \caption{{Basis vectors for irreducible representations in the 12-dimensional space of translationally-invariant dimer coverings. Here $\omega = \exp \left(2 \pi i / 3\right)$. The inversion eigenvalue $\lambda = +1$ corresponds to $\mathbf{g}$--representations and $\lambda = -1$ corresponds to $\mathbf{u}$--representations. The $E_{\pm}^{\mathbf{g/u}}$ basis states span the 2-dimensional representation and $T_{a/b/c}^{\mathbf{g/u}}$ basis states span the 3-dimensional representation.}}
    \label{tab:dimer_irreps}
\end{table}

\section{Dimer correlation scaling}
\label{sec:appendix_scaling}
In order to justify the $A + B \Omega^{-1}$ scaling used in Fig.~\ref{fig:susceptibilities} to extrapolate dimer-dimer susceptibility correlations to thermodynamic limit in dimerized phase, we employ the approach of Ref.~\cite{PhysRevB.84.024406} (see Fig. 14 therein) and of Ref.~\cite{cite_2}. Unlike magnetic order, which breaks continuous $SU(2)$ symmetry and leads to emergence of gapless Goldstone modes, dimer order breaks no continuous symmetry and we expect that all excitations are gapped. As the result, dimer-dimer correlations saturate with distance exponentially fast $\langle \hat D_i \hat D_j \rangle \sim A + B \exp(-r / \xi).$ This fast saturation leads to the inverse-volume finite-size correction scaling\footnote{Strictly speaking, following this logic, the $\Omega^{-1}$ extrapolation used for inversion-breaking susceptibility in the magnetic phase at $j_2 / j_1 = 0.2$ in Fig.~\ref{fig:susceptibilities} should not be used in this case, as we assume no dimer order establishment. In absence of dimer order, dimer-dimer correlations will saturate polynomially and will lead to $L^{-\alpha}$ scaling with $\alpha < 3$. This, however, will not change the conclusion since a steeper slope will only lead to an even smaller extrapolation result.}. Following Ref.~\cite{PhysRevB.84.024406}, in Fig.~\ref{fig:dimer_r_4x4x4} we show $\langle \hat D_0^{01} \hat D_j^{\alpha} \rangle$ - $\langle \hat D_0^{01}\rangle \langle \hat D_j^{\alpha} \rangle$, where $\alpha$ labels the 12 bonds that belong to a unit cell as introduced in Sec.~\ref{subsec:general_analysis} and in Fig.~\ref{fig:dimerisation_ed}, as the function of distance $|\mathbf r_0 - \mathbf r_j|$ between the unit cells $0$ and $j$, measured in the non-magnetic phase at the $4 \times 4^3$ cluster within mVMC. Notably, the correlations quickly saturate to non-vanishing values, which are dependent on $\alpha$. In another words, we expect small correlation length $\xi \ll L$, which paves way to finite dimer order parameter susceptibility and justifies the scaling employed in Fig.~\ref{fig:susceptibilities}.

\begin{figure}[t!]
    \centering
     \includegraphics[width=0.5\textwidth]{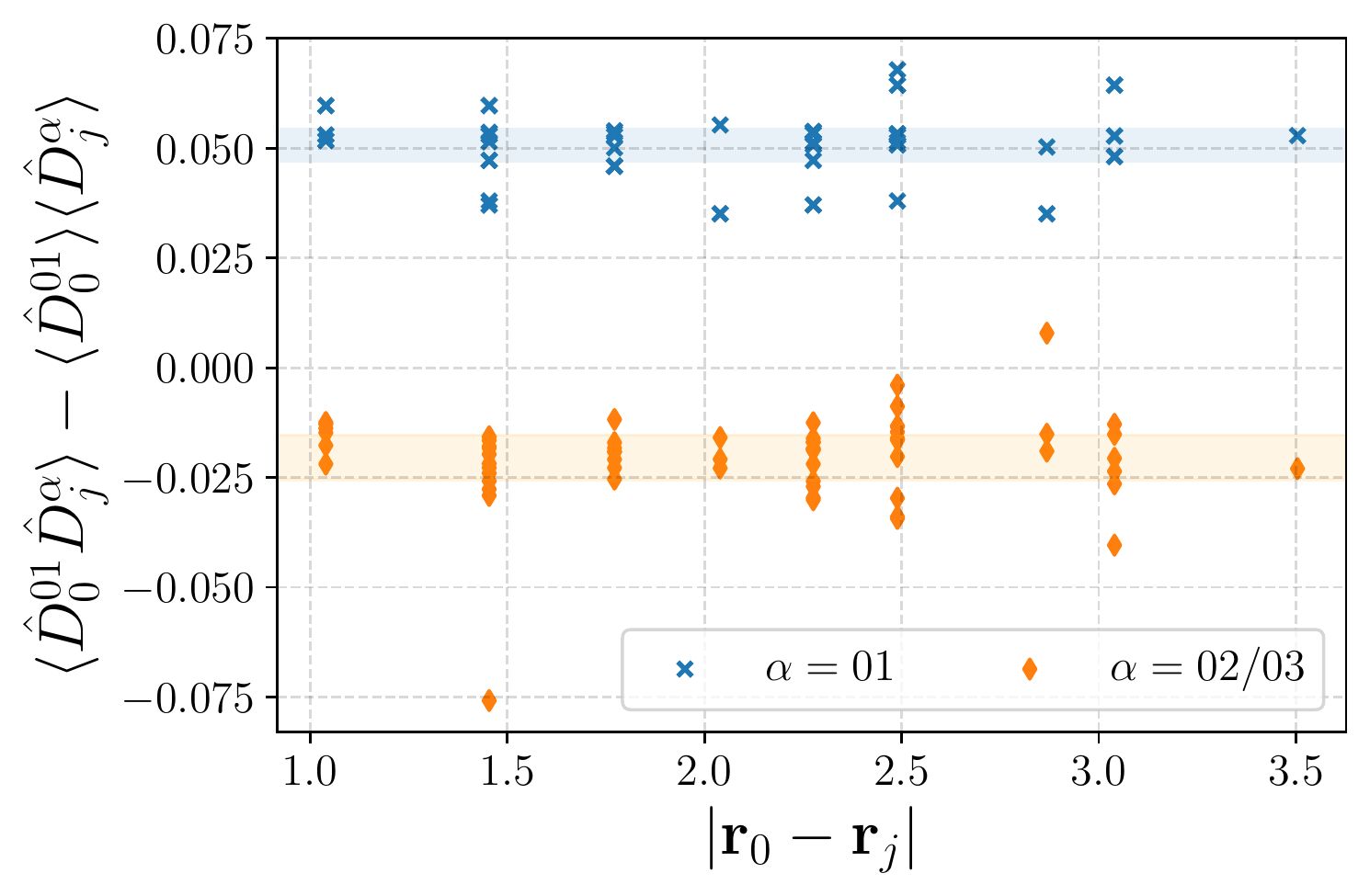}
    \caption{{Bond-bond dimer correlations $\langle \hat D_0^{01} \hat D_j^{\alpha} \rangle$ - $\langle \hat D_0^{01}\rangle \langle \hat D_j^{\alpha} \rangle$ resolved in real space and measured on the $4 \times 4^3$ cluster within $SU(2)$ mVMC at $j_2 / j_1 = 0$. The distance $|\mathbf r_0 - \mathbf r_j|$ is between the origins of the unit cells $0$ and $j$, index $\alpha$ runs over the 12 bonds in a unit cell as introduced in Sec.~\ref{subsec:general_analysis} and in Fig.~\ref{fig:dimerisation_ed}. Shaded regions indicate mean and standard deviation of the measured correlations. The point $\mathbf r_0 = \mathbf r_j$ is omitted due to big difference with the other values.}}
    \label{fig:dimer_r_4x4x4}
\end{figure}
}

\begin{table}[b!]
    \centering
    \begin{tabular}{|c|c|c|c|c|}
    \hline
     geometry & mVMC & NQS CNN & ED & DMRG~\cite{hagymasi2020possible}  \\ \hline
     $4 \times 2^3$ & -0.5162(1) & -0.5128(1) & -0.5168 & -0.5168 \\
     $ $ & 0.7275(1) & 0.766(3) & 0.6872 & 0.6872 \\ \hline
     $4 \times 2^2 \times 3$ & -0.5028(2) & -- & -- & -0.5029  \\ 
     $ $ & -- & -- & -- & 0.36(3) \\ \hline
     $4 \times 2^2 \times 4$ & -0.4924(2) & -0.4908(2) & -- & 0.4923 \\
     $ $ & 0.36(5) & -- & -- & -- \\ \hline
     $4 \times 2^2 \times 6$ & -0.4917(4) & -0.490(1) & -- & -- \\
     $ $ & 0.446(5) & -- & -- & -- \\ \hline
     $4 \times 3^3$ & -0.4871(1) & -0.4826(2) & -- & -0.4851 \\
     $ $ & 0.625(5) & 0.65(4) & -- & -- \\ \hline
     $4 \times 2 \times 4 \times 6$ & -0.4915(2) & -- & -- & -- \\
     $ $ & 0.463(1) & -- & -- & -- \\ \hline
     $4 \times 4^3$ & -0.4831(1) & -- & -- & -- \\
     $ $ & 0.55(1) & -- & -- & -- \\ \hline
     $12$~\cite{PhysRevLett.80.2933} & -- & -- & -- & -- \\
     $ $ & -- & -- & 0.7 & -- \\ \hline
     $12'$~\cite{doi:10.1143/JPSJ.70.640} & -- & -- & -- & -- \\
     $ $ & -- & -- & 0.5491 & -- \\ \hline
     $28$~\cite{PhysRevB.97.144407} & -- & -- & -0.4820 & -- \\
     $ $ & -- & -- & 0.201 & -- \\ \hline
     $36$~\cite{PhysRevB.97.144407} & -- & -- & -0.4669 & -- \\
     $ $ & -- & -- & 0.0832 & -- \\ \hline
    \end{tabular}
    \caption{Energy and singlet-triplet gap obtained in this work and other papers, if the reference number is given. The ground state energy per site is listed in the first line of every cell, while second line contains the absolute spin singlet-triplet gap magnitude, not normalized to the system volume.}
    \label{tab:energies_spingaps}
\end{table}

\section{Ground state energies and spin gap}
\label{sec:appendix_table}
The energy-per-spin is the foremost way to assess the quality of the variational wave function. The spin-1/2 pyrochlore ground state energy at $j_2 / j_1 = 0$ was previously estimated within various approaches~\cite{PhysRevB.79.144432,PhysRevB.101.174426,PhysRevB.100.024424,Kim_2008,PhysRevB.63.144432,PhysRevB.61.1149}. In Table\,\ref{tab:energies_spingaps} we summarize the ground state energies from this work as well as other references which employ DMRG or exact diagonalization. Note that the quasi-planar geometry used in exact diagonalization of $28$- and $36$-spin clusters in~\cite{PhysRevB.97.144407} has a significant effect on the ground state energy, which emphasizes strong geometric dependence of the results and the importance of choosing clusters with the maximum possible number of spatial symmetries.

Similarly, the non-vanishing spin gap is important not only for proving the absence of magnetic order, but is also essential for the perturbative dimer model analysis of the pyrochlore~\cite{doi:10.1143/JPSJ.70.640,PhysRevB.65.024415}. In Table\,\ref{tab:energies_spingaps} we also show the spin gap obtained in this work and other studies on specific clusters. Note again the strong dependence of spin gap on the cluster geometry. 
\bibliography{refs}
\end{document}